\documentclass[epjST]{svjour}
\usepackage{amsmath,bm,graphicx}
\usepackage{amssymb}
\usepackage{amsfonts}
\usepackage{ulem}
\begin{document}
\title{Working under confinement}
\author{Paolo Malgaretti\inst{1,2}\fnmsep\thanks{\email{malgaretti@is.mpg.de}} \and Ignacio Pagonabarraga\inst{3} \and J.M. Rubi\inst{3}}
\institute{Institut f\"{u}r Intelligente Systeme, Heisenbergstr. 3 D-70569 Stuttgart Germany \and IV. Institut f\"{u}r  Theoretische Physik, Universit\"{a}t Stuttgart, Pfaffenwaldring 57, D-70569 Stuttgart, Germany \and Departament de Fisica Fonamental, Universitat de Barcelona, C. Mart\'{i} i Franques 1, Barcelona, Spain}
\abstract{We analyze the performance of a Brownian ratchet in the presence of geometrical constraints. A two-state model that describes the kinetics of molecular motors is used to characterize the energetic cost when the motor proceeds under confinement, in the presence of an external force. We show that the presence of geometrical constraints has a strong effect on the performance of the motor. In particular, we show that it is possible to enhance the ratchet performance by a proper tuning of the parameters characterizing the environment. These results open the possibility of engineering entropically-optimized transport devices.
} 
\maketitle
\section*{Introduction}
\label{intro}

Transport of particles at small scales is frequently mediated by the presence of obstacles that particles may find along their trajectories. The effect of the obstacles is to diminish the space accessible to the particles. Since the number of microstates is proportional to the volume of the accessible space, the entropy varies along the trajectories and the transport proceeds through an entropic potential landscape. This transport referred to as entropic transport~\cite{jacobs,zwanzig,Reguera2001,Reguera2006} exhibits peculiar characteristics, very different from those for the case of unbounded systems~\cite{Vazquez,Bezrukov,burada2009,malgarettiFrontiers}. The case of transport of particles in narrow and tortuous channels where the presence of obstacles and irregularities of the boundaries alter their trajectories is a typical example. Motion of particles in the interior of a living cell~\cite{Witz2009} or through an ion channel~\cite{Hille,ion_channel_faraudo}, diffusion in zeolites~\cite{zeolites} and in 
microfluidic devices~\cite{Bezrukov,Han,Fujita,Malgaretti2014PRL}, and folding of proteins modeled as motion of the state of the protein through a phase space funnel-like region~\cite{Dill} are cases in which the system proceeds in a bounded region.  Confinement is a source of continuous dynamic changes and consequently of modifications of the global properties of a system.

A peculiar scenario arises when Brownian particles diffuse in a ratchet potential. The lack of a detailed balance principle  gives rise to motion of the particles, a situation not found when the particles are subjected to constant forces.  Brownian ratchets show different dynamical behaviours for single particles~\cite{marchesoniRMP} and for particle collective behavior~\cite{guerin,Malgaretti2012PRL} in an homogeneous medium.
Interestingly, it has been shown that the interplay between the ratchet mechanism and the geometrical constraint can give rise to rectification even when the ratchet alone would not do so. Such a novel dynamical scenario, namely cooperative rectification~\cite{Malgaretti2012}, has been shown to produce net currents, even in the case of a symmetric ratchet potential and even current inversion, i.e. a particle moving against the direction an asymmetric ratchet per se would provide~\cite{Malgaretti2012,Malgaretti2013}. 

Even though cooperative rectification has been shown to give rise to intriguing dynamic scenarios, up to now the question about its performance, i.e. the energy consumption per unit displacement of cooperative rectification has not been yet addressed. Here we shall discuss whether cooperative rectification is costly when compared to the bare ratchet mechanism or if the presence of entropic barriers can enhance the overall performance of the Brownian ratchet. In particular, we will address two different scenarios, namely, the case in which the Brownian ratchet displaces against a drag force and the case in which, on the top of the drag force, Brownian ratchets are also pulling against a conservative force. While in the latter case a good definition of the performance is provided by the thermodynamic efficiency, in the former case such a definition is lacking and therefore we will rely on an ad hoc observable that allows us to capture the energy cost per unit displacement of the molecular motor. 

\section*{Confined Brownian ratchets}

Brownian ratchets have been widely used as models to understand how molecular~\cite{Guerin2010,Astumian} as well artificial~\cite{Allison2002,Zhu2004,Linke1999} motors operate. To analyze their performance under confinement, we shall consider that particles move in a periodic channel whose cross-sectional area varies along the $x$-direction and is constant along $z$. The space available to the center of mass of a particle with radius $R$ is $2 h(x)L_z$, where $L_z$ is the width along the $z$-direction and
\begin{equation}
 h(x)=h_0-R+h_1 \sin \left[\frac{2\pi}{L}(x+\Delta\phi)\right]
\label{channel}
\end{equation}
being $h(x)+R$ the half-width of the channel along the $y$-direction and $\Delta \phi$ the phase shift between the periodicity of the channel and of the ratchet potential.

As a ratchet model, we use the two-state model~\cite{Julicher1997} that provides a simple framework to describe  molecular motor motion. According to this model, a Brownian particle jumps between two states,  $i=1,2$, (strongly and  weakly bound) that determine under which potential, $V_{i=1,2}$, it displaces~\cite{Julicher1997}.  A choice of the jumping rates $\omega_{12,21}$ that break detailed balance, together with an asymmetric potential  of the bound sate, $V_{1}(x)$, determines the average molecular motor velocity $v_0\neq0$.  The conformational changes of the molecular motors  introduce an additional  scale that competes with  rectification and geometrical confinement.  Infinitely-processive molecular motors remain always attached to the filament along which they move and are affected by the geometrical restrictions only while displacing along the filament; accordingly, we choose  channel-independent binding rates $\omega_{21,p}(x) = k_{21}$. On the contrary, highly non-processive 
molecular 
motors detach frequently  from the  biofilament  and diffuse away; this effect is accounted for a channel-driven binding rate, $\omega_{21,np}(x)=k_{21}/h(x)$. Motors jump to the weakly bound state only in a region of width $\delta$ around the  minima of  $V_1(x)$,  with rate $\omega_{12}  =   k_{12}$. Accordingly, the motor densities in the strong(weak) states, $P_{1(2)}$ along the channel follow~\cite{Julicher1997}
\begin{equation}\label{FJ-two-states}
\begin{array}{cc}
\partial_{t}P_{1,2}({\bf r})+\nabla\cdot \mathbf{J}_{1,2}  = \mp \left[\omega_{12}(x)P_{1}({\bf r})+\omega_{21}(x)P_{2}({\bf r})\right]\\
\end{array}
\end{equation}
where 
\begin{equation}
\mathbf{J}_{1,2}({\bf r}) =  -D\big[{\bf  \nabla} P_{1,2}({\bf r})+P_{1,2}({\bf r}) {\bf \nabla} \beta U_{1,2}({\bf r})\big] 
\label{eq:currents}
\end{equation}
stands for the current densities in each of the two states in which motor displaces, $\beta = 1/(k_B T)$  corresponds to the inverse thermal energy for a system at temperature  $T$, $k_B$ stands for Boltzmann's constant, and $U_{1,2}$, which is periodic,$ U_{1,2}(x,y,z)=U_{1,2}(x+L,y,z)$,  reads
\begin{equation}
U_{1,2}(x,y,z) =\left\{
\begin{array}{cc}
V_{1,2}(x) &|y|\le h(x)\, \&\, |z| \le L_z/2 \\
\infty & |y|> h(x)\, or\, |z|>L_z/2
\end{array}
\right.
\label{potential}
\end{equation}
where 
\begin{equation}
V_{1,2}(x) =\left\{
\begin{array}{cc}
 \frac{\Delta V}{2} \sin  \frac{2 \pi}{L} x  & \mbox{bound state}\\ \frac{\Delta V}{2} & \mbox{otherwise}
\end{array}
\right.
\label{eq:V0}
\end{equation}
is the potential stemming from the interaction of the motor with the filament characterized by a depth $\Delta V$. 

We shall assume that motor distribution equilibrates much faster in the cross section of the channel than along it. It is then  possible to project the $3D$ convection diffusion equation onto an effective $1D$ equation governing the dynamics of particle density along the longitudinal direction. Such a procedure was first introduced by Jacobs~\cite{jacobs} and subsequently improved by Zwanzig~\cite{zwanzig} and by ~\cite{Reguera2001,Kalinay2005,Kalinay2006,martens2011} and tested~\cite{burada2007} in a variety of scenarios~\cite{Reguera2006,Reguera2012,Dagdug2012,Kalinay2014} 

This regime is fulfilled for a smoothly varying-section channels, $\partial_x h \ll 1$, in which the particle distribution reaches local equilibrium in the cross section almost immediately. We can then  approximate the  profile of the probability distribution function, $P(x,y,t)$,   assuming it  equilibrates  in the cross section of the channel, and write 
\begin{equation}
P_{1,2}(x,y,z,t)  =  \mathcal{P}_{1,2}(x,t)\frac{e^{-\beta V_{1,2}(x,y)}}{e^{-\beta A_{1,2}(x)}}
\end{equation}
where 
\begin{equation}
e^{-\beta A_{1,2}(x)}  =  \int_{-L_z/2}^{L_z/2}\int_{-h(x)}^{h(x)}e^{-\beta V_{1,2}(x,y)}dy dz 
\label{free-en}
\end{equation}
$A_{1,2}(x)$ corresponding to the potential of the mean force that can be identified with the free energy in the coarse-grained description. Depending on the motor internal state, two  free energies,  $A_{1,2}(x)=V_{1,2}(x)-k_BT S_{1,2}(x)$, account for the interplay between the biofilament interaction and the channel constraints. 

By integration of Eq.~\ref{FJ-two-states} over $dy,dz$ we then obtain
\begin{equation}
 \dot {\mathcal{P}}_{1,2}(x,t)=\partial_x D\left[ \beta \mathcal{P}_{1,2}(x,t)\partial_x A_{1,2}(x)+\partial_x \mathcal{P}_{1,2}(x,t)\right],
\label{FJ1}
\end{equation}
which encodes both the confining as well conservative potentials given by Eq.~\ref{potential} in the free energy  $A_{1,2}(x)$.
Since all quantities of interest are independent of $z$, without loss of generality we can assume $\int_{-L_z/2}^{L_z/2}dz=1$. Defining the average, $x$-dependent, contribution to the  energy coming from conservative potentials as
\begin{equation}
 W_{1,2}(x)=e^{\beta A_{1,2}(x)}\int_{-h(x)}^{h(x)}V_{1,2}(x,y)e^{-\beta V_{1,2}(x,y)}dy,
\label{avg-V}
\end{equation}
 we can define the entropy along the channel as $TS_{1,2}(x)=W_{1,2}(x) -A_{1,2}(x)$ using Eq.(~\ref{free-en}), leading to  
\begin{equation}
 S_{1,2}(x)=\ln\left(\int_{-h(x)}^{h(x)}e^{-\beta V_{1,2}(x,y)}dy\right)+\beta W_{1,2}(x).
\label{entropy}
\end{equation}
In the linear regime, reached when  $\beta V_{1,2}(x,y) \ll 1$, Eq.~\ref{entropy} yields:
\begin{equation}
 S_{1,2}(x)\simeq \ln(2 h(x))
\end{equation}
where entropy has a clear geometric interpretation, being the logarithm of the space, $2h(x)$, accessible to the center of mass of the tracer. Accordingly,  we introduce the entropy barrier, $\Delta S$, defined as
\begin{equation}
 \Delta S_{1,2}=\Delta S=\ln \left(\frac{h_{max}}{h_{min}}\right),
\label{entropy-barrier}
\end{equation} 
which represents the difference in the entropic potential  evaluated at the maximum, $h_{max}$, and minimum, $h_{min}$  channel aperture. Eq.~\ref{entropy-barrier}, together with Eq.~\ref{channel}, shows that the geometrical information coming from both the channel and the particle are encoded in $\Delta S$.

\section*{Effective performance}
When molecular motors move in a crowded environment, their stepping performances are strongly affected by the geometrical constraints~\cite{Malgaretti2013}. A question that has not been addressed up to now is how the energy consumed by the motor is affected by confinement.
\subsubsection*{Absence of external forces}
When the motor does not displace against an applied external force, the standard definition of efficiency\footnote{Here we refer to the efficiency as obtained from the first law of thermodynamics}, namely the mechanical work performed over the energy consumed is not adequate since the mechanical work performed vanishes. In this case, it is more convenient to compute the amount of energy consumed to perform one step forward~\cite{Suzuki2003,Machura,Malgaretti2012PRL,Rubi2012}. 

The two state model is particularly insightful because it clearly identifies the mechanism  where energy is consumed. Then, an unambiguous  definition of the energy consumption per unit step  can be defined. Each jump of the molecular motor between the bound  and the weakly bound state costs an energy, $\Delta V$, see Eq.~\ref{eq:V0}, equal to  the  ratchet potential depth. Therefore knowing the average number of jumps needed by a motor to perform one step froward, we can calculate the energy consumption per unit step. From Eq.~\ref{FJ-two-states}, we can introduce $\Omega$,  the average number of jumps a motor performs per unit time  between the strongly and weakly bound states,
\[
\Omega=\int_{0}^{L}w_{off}(x)p_{1}(x)dx=\int_{0}^{L}w_{on}(x)p_{2}(x)dx
\]
Having obtained this quantity, we can define the average number of jumps between the two states needed to perform a single step as 
\begin{equation}
\Gamma=\frac{\Omega L}{v}\label{eq:gamma}
\end{equation}
where $v$ is the average motor velocity. $\Gamma$, a  dimensionless parameter, is proportional to the energy consumed by a molecular motor to perform a single step. 

\begin{figure}
\includegraphics[scale=0.34]{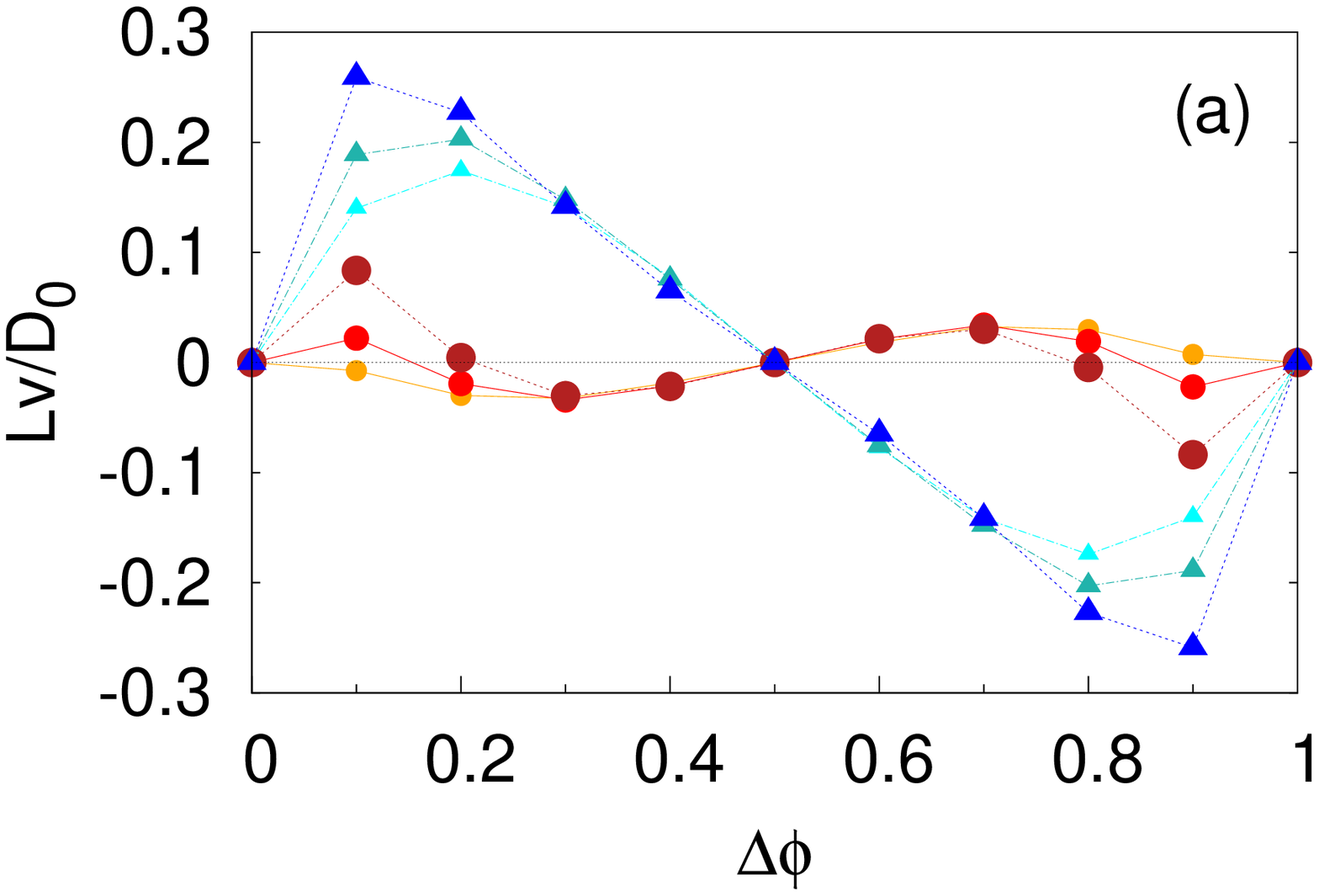}\includegraphics[scale=0.34]{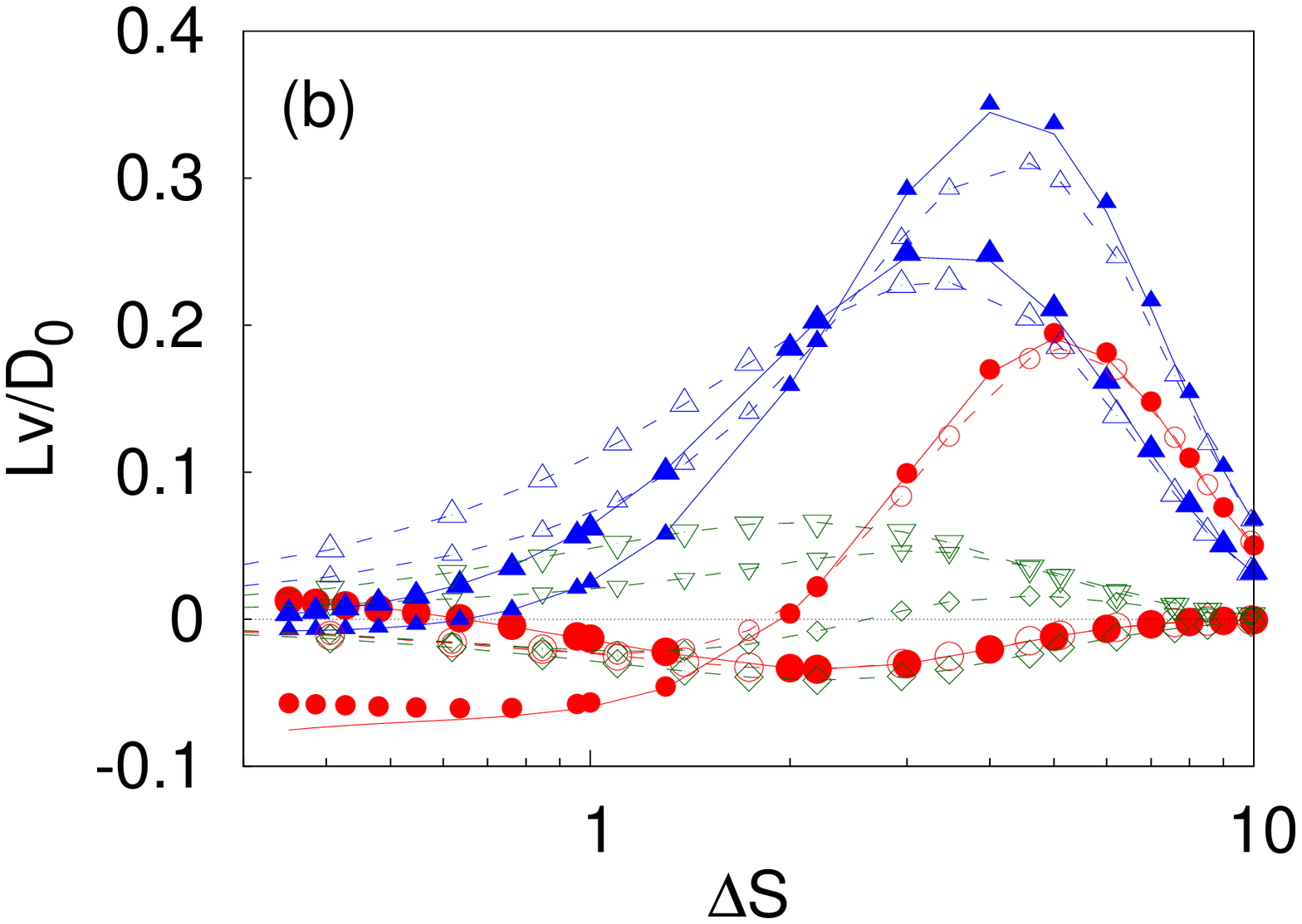}
\caption{Rectification of a processive (circles), non-processive (triangles) Brownian particle moving due to the two state model in a symmetric channel. (a): particle velocity, in units of $D_{0}/L$, with $D_0=D_0(R=1)$, as a function of the phase shift $\Delta\phi$  for different values of the parameter $\Delta S=1.73,2.19,2.94$ (the larger the symbol   size, the larger $\Delta S$), for  $\Delta V=0.4$ and $\omega_{2,1}/\omega_{1,2}=0.01$. (b): Processive (circles), non-processive (triangles) Brownian motor velocity, in units of $D_{0}/L$, as a function of $\Delta S$ upon variation of particle radius $R$ (solid lines, for $h_0=1.25,h_1=0.2$), $h_0$ (solid points, for $R=1,h_1=0.2$) or $h_1$ (open points, for $R=1,h_0=1.25$) for $\Delta\phi=0.1,0.2$(larger symbols correspond to larger  $\Delta\phi$). Reprinted with
permission from Malgaretti, Pagonabarraga and Rubi J. Chem. Phys. 138, 194906 (2013). Copyright (2013) by the American Physical Society.}
\label{fig-twostate-simm-simm}
\end{figure}

\subsubsection*{Kinetics under external forces}
When molecular motors are stepping against an externally applied force, it is possible to define a proper efficiency as the ratio between the work performed against the conservative force, $|f|L$ over the energy consumed, $\Delta V \Gamma$, when the motor displaces a single period of the potential, $L$. Therefore we can define the efficiency as 
\begin{equation}
\eta=\frac{|f|L}{\Delta V\Gamma}=\frac{|f|v}{\Delta V \Omega}
\label{efficiency}
\end{equation}
that can be easily recognized as the power transformed in mechanical work over the power injected that allows the molecular motor  to jump between its two internal states.

\section*{Cooperative rectification}
 
In order to analyze the interplay between the ratchet potential and the geometrical constraints, in the following we assume that $L=1$. The most striking feature of that interplay takes place when both  the ratchet potential and the channel shape are symmetric~\cite{Malgaretti2012}. Despite the fact that in this case none of the mechanisms can rectify independently, rectification can be observed~\cite{Malgaretti2012,Malgaretti2013}.


Fig.~\ref{fig-twostate-simm-simm}.a  shows that, even though the ratchet potential and the channel are both symmetric along the longitudinal direction, a  net particle current develops. In particular, such a current sets when the ratchet potential and the channel corrugation are out of registry.
Cooperative rectification is quite sensitive to the entropic barrier $\Delta S$. As shown in Fig.~\ref{fig-twostate-simm-simm}.b the net current can be modulated and its sign can be inverted by tuning the amplitude of $\Delta S$ at least for processive motors\footnote{Similar behavior has been obtained for non processive motors when the ratchet potential is asymmetric, as shown in reference~\cite{Malgaretti2012,Malgaretti2013}}. 

\section*{Working under confinement}

\subsection*{No external force}

\begin{figure}
\includegraphics[scale=0.33]{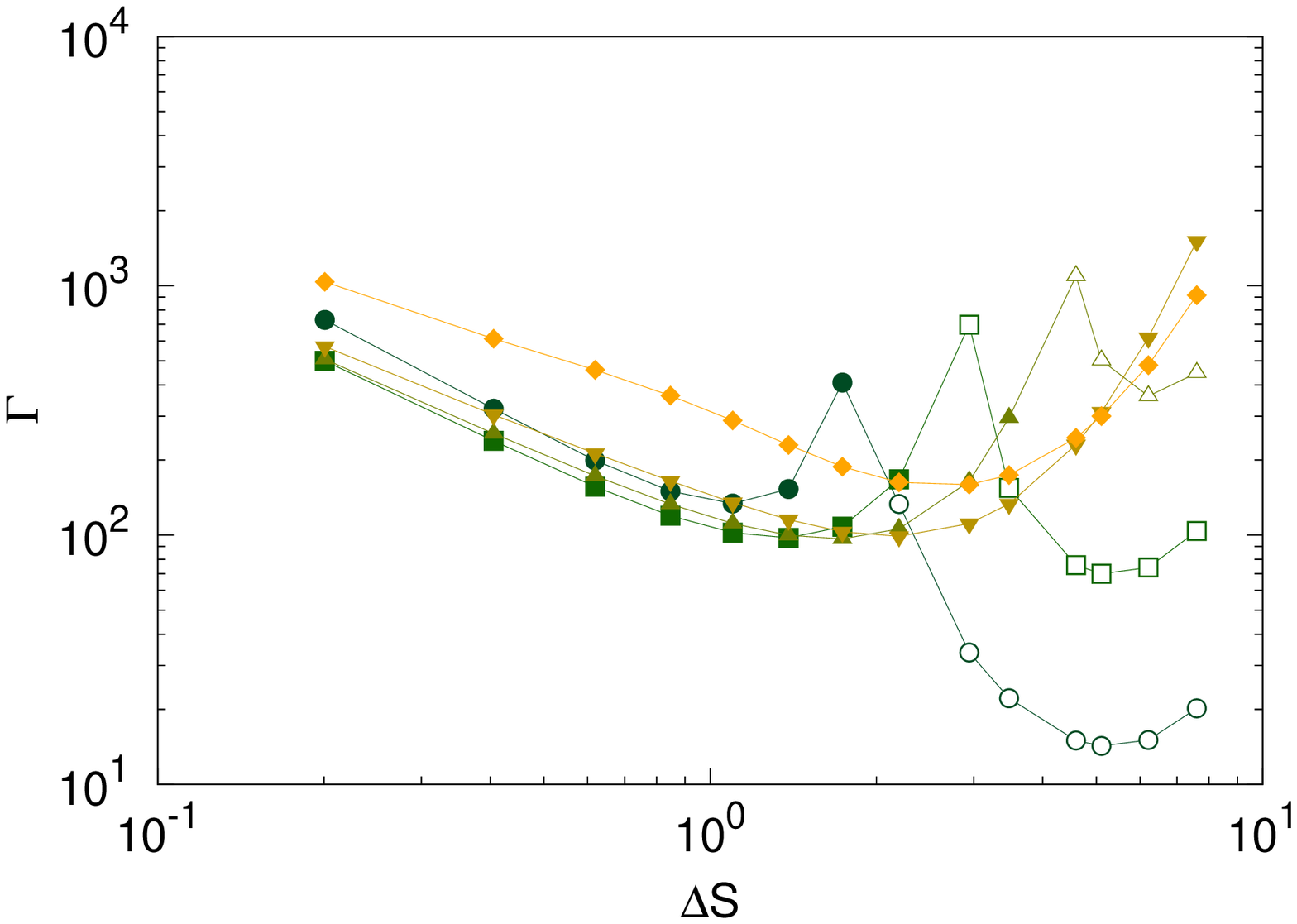} \includegraphics[scale=0.33]{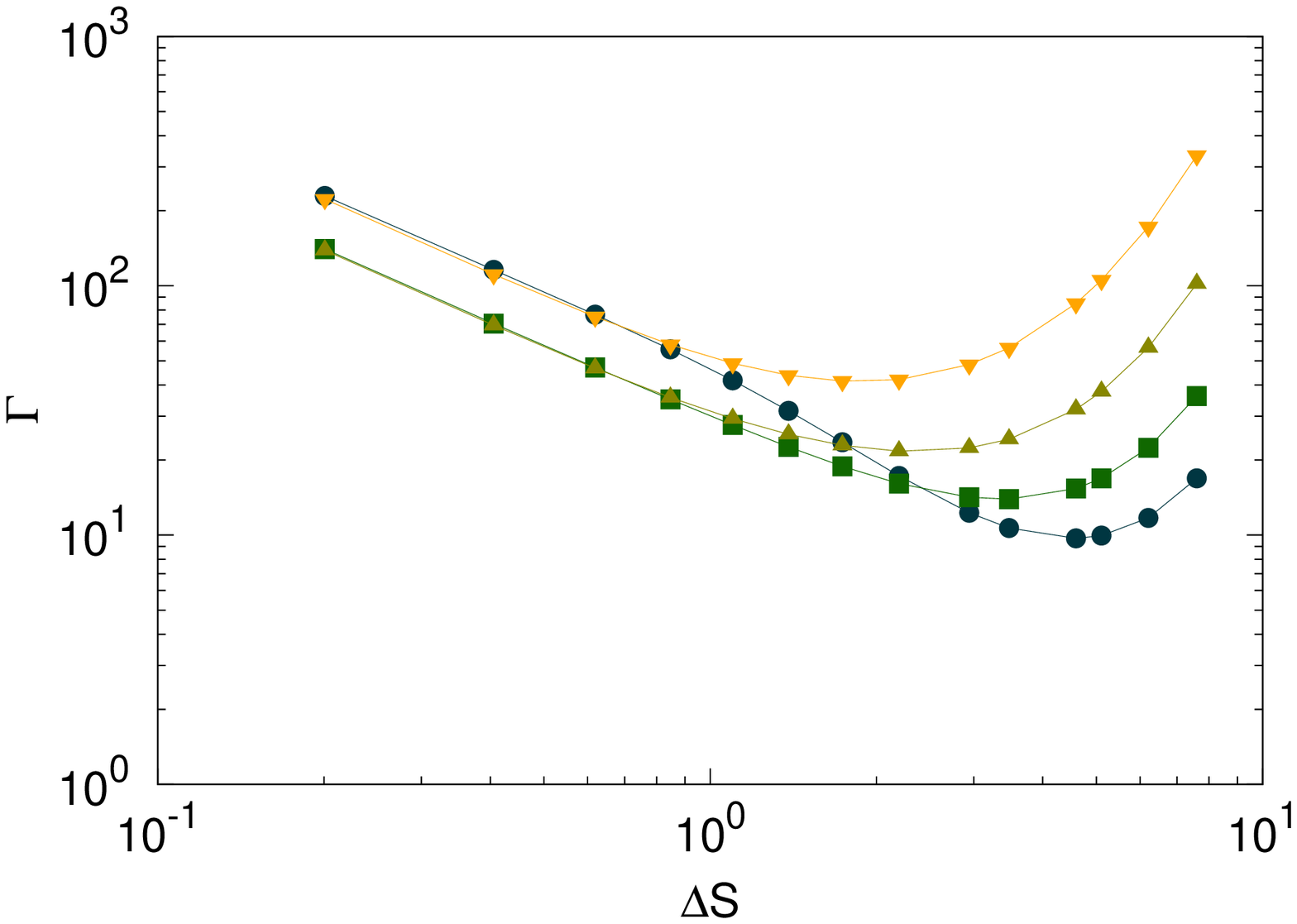}\\
\caption{$\Gamma$ as a function of the entropic barrier $\Delta S$ for different
values of the phase shift where lighter lines stands for larger values
of $\Delta\phi$ for a processive motor (left, $\Delta\phi=0.1,0.2,0.25,0.3,0.4$
) and non processive motor (right, $\Delta\phi=0.1,0.2,0.3,0.4$) in the
case of a symmetric ratchet potential, characterized by $\beta \Delta V=10$, and symmetric channel shape.
In the case of processive motors velocity changes sign upon increase
of $\Delta S$ as marked in the figure: filled (open) points stands
for negative (positive) velocities. \label{fig:simm-simm-no-force}}
\end{figure}

We will first analyze the motion of a motor in the absence of an applied external force in a symmetric channel, and will analyze the efficiency of motor transport for both a symmetric and an asymmetric ratchet potential. In the former case, cooperative rectification emerges  when  the phase
shift between the ratchet and the channel are not in register. In the latter scenario we will assess  how  the modulation of the net motor motion due to the geometrical
constraints affects its efficiency.  

\begin{figure}
\includegraphics[scale=0.33]{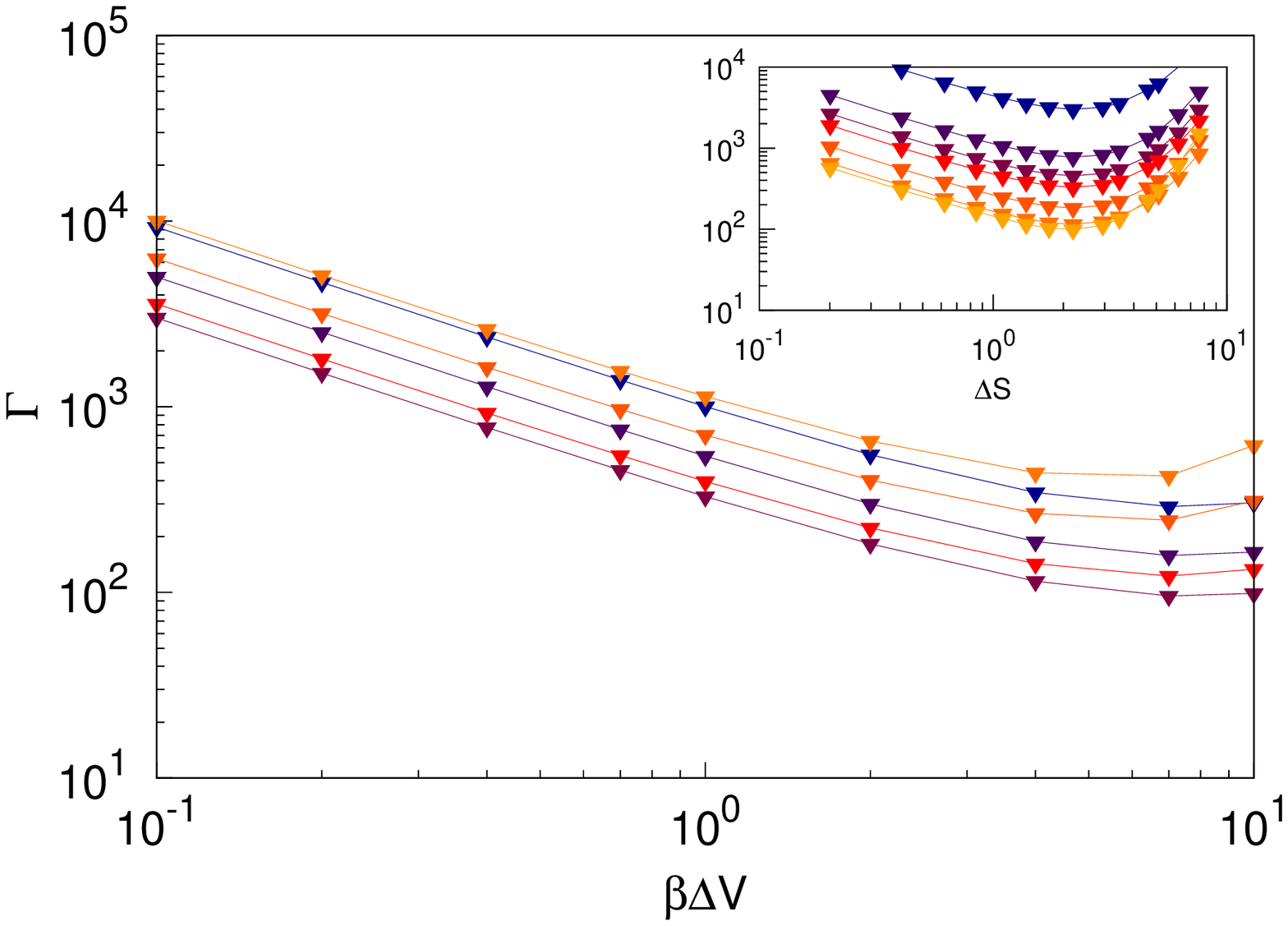} \includegraphics[scale=0.33]{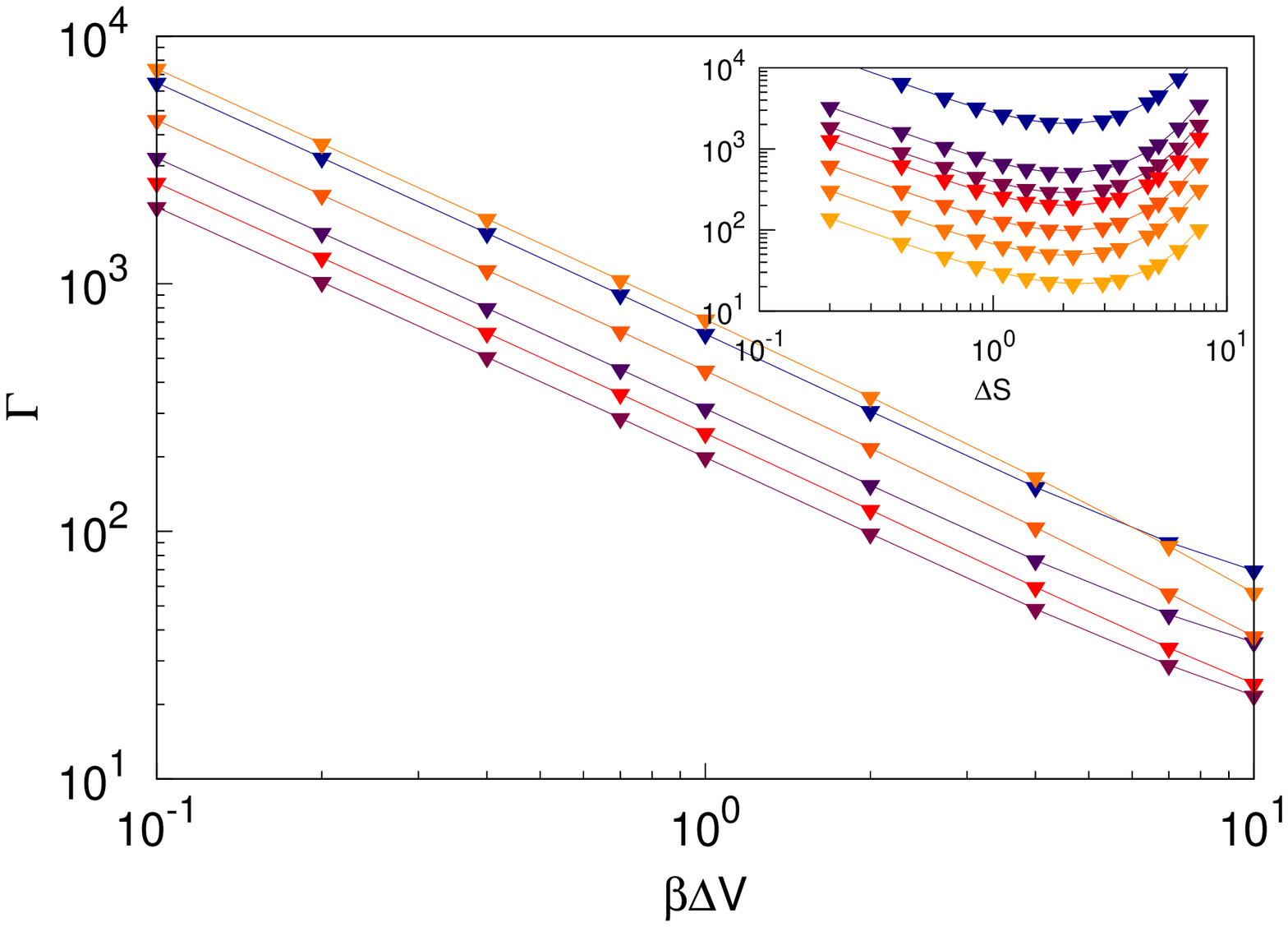}
\caption{ $\Gamma$ as a function of the ratchet potential depth, $\Delta V$, for  processive  (left) and non processive  (right) symmetric molecular motors  moving along a  symmetric channel, for a phase shift $\Delta\phi=0.3$. The different curves correspond to different degrees of confinement, quantified by the magnitude of the 
 entropic barrier, $\Delta S=0.2,0.85,2.2,3.5,5.1,7.6$; the lighter the color of the curve the larger the value of $\Delta S$. Inset: $\Gamma$ as a function of $\Delta S$ for $\beta \Delta V=0.1,0.4,0.7,1,4,7,10$; the lighter the color of the curve the larger the value of  $\delta V$.
\label{fig:simm-simm-no-force-velVspot}}
\end{figure}

\subsubsection*{Symmetric channel and ratchet potential}

In this case, we expect cooperative rectification to generate net fluxes
when the ratchet potential and the channel are not in registry. Fig.~\ref{fig:simm-simm-no-force} shows the dependence of $\Gamma$ on
the entropic barrier $\Delta S$ for different values of the phase
shift. Both processive and non processive motors show a non monotonic
behavior of $\Gamma$ as a function of $\Delta S$ therefore identifying
an optimal value of $\Delta S$ for which $\Gamma$ is minimum. Hence 
the spatial constriction induced by the  channel plays a double role. On one hand geometric constraints, jointly with the motor mechanism, are responsible for the onset of cooperative rectification and therefore for the net current of motors. On the other hand channel corrugation affects the energetic cost of displacing along a in a corrugated environment. Overall,  increasing channel corrugation does not necessary lead to an increase
in the energy consumed per unit step, as shown in Fig.\ref{fig:simm-simm-no-force}.
For a finite set of values of $\Delta\phi$, processive motors change their displacement direction upon increasing $\Delta S$~\cite{Malgaretti2012}, see Fig.~\ref{fig-twostate-simm-simm}. Accordingly,  $\Gamma$ shows a second divergence at the corresponding $\Delta S$ value\footnote{$\Gamma$ diverges when the motor spends energy  but is not displacing along the channel.}. For the parameters analyzed in the Figure,  non processive motors do not change their directionality  and $\Gamma$  only diverges for a  uniform channel, $\Delta S=0$, when the motor does not rectify.
Fig.~\ref{fig:simm-simm-no-force-velVspot} shows the dependence of $\Gamma$ on the depth of the ratchet potential $\Delta V$. Increasing $\Delta V$ decreases the value of $\Gamma$ till a plateau is reached for $\beta\Delta V\simeq 10$. Interestingly, we have found that the value of $\Delta S$ that minimizes $\Gamma$ is quite robust under variation of the ratchet potential depth, as shown in the insets of Fig.~\ref{fig:simm-simm-no-force-velVspot}.

\begin{figure}
\includegraphics[scale=0.33]{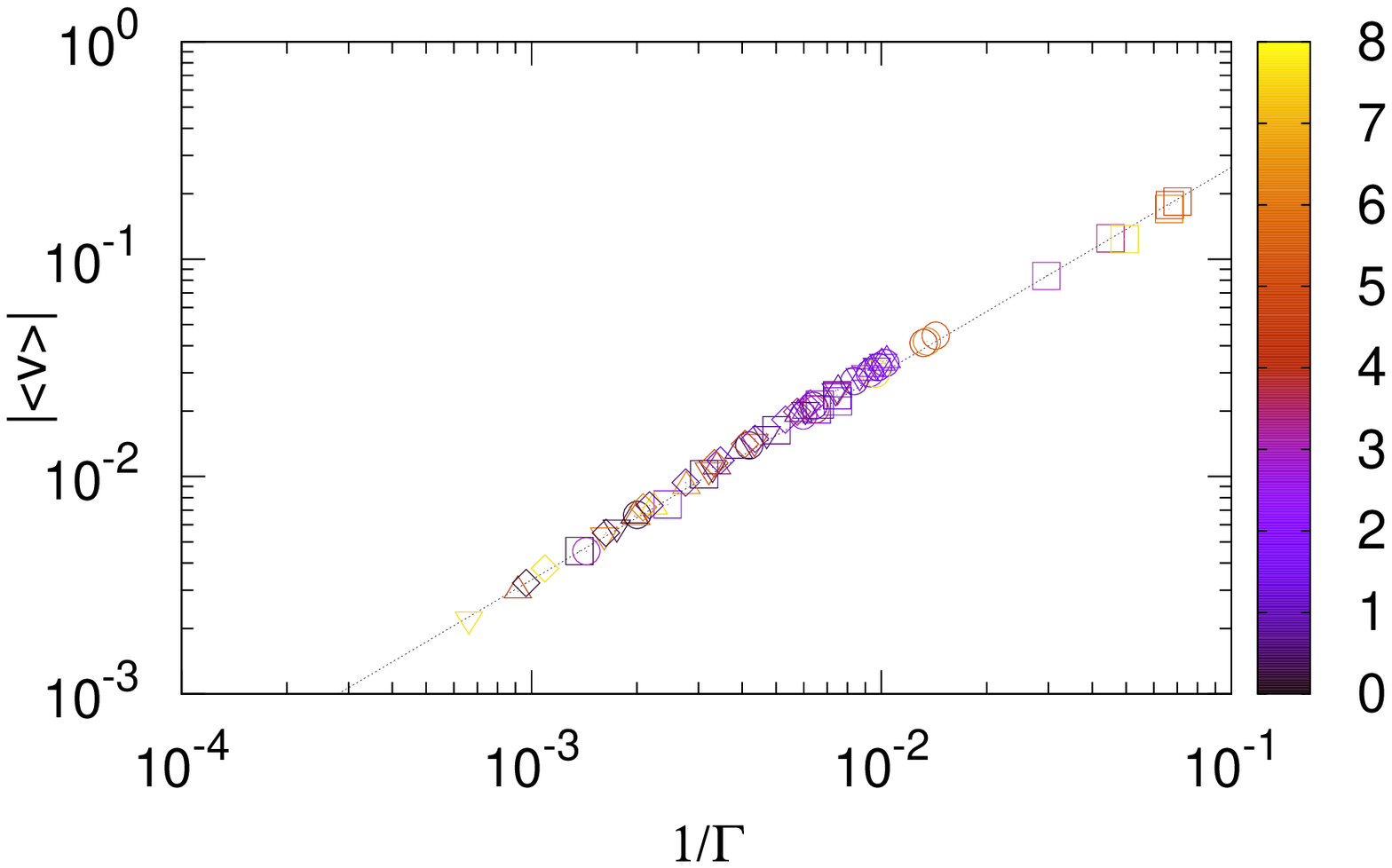} \includegraphics[scale=0.33]{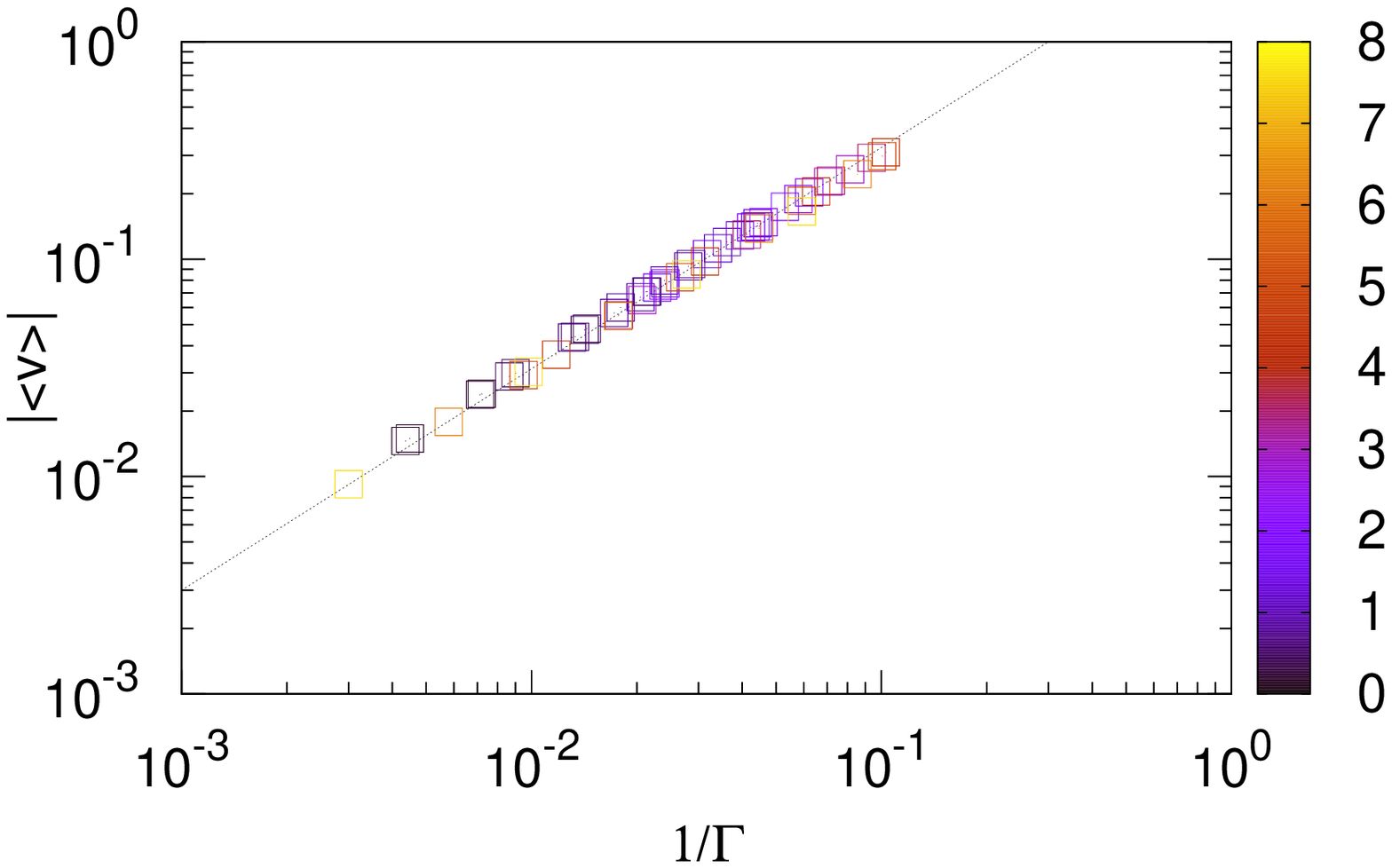}
\caption{Mean motor velocity, $v$,  of  processive (left) and non processive (right) symmetric molecular motors  moving along a  symmetric channel  as a function of $\Gamma$. Different symbols  correspond to different phase shift, quantified by $\Delta \phi=0.1$ (squares), $\Delta \phi=0.2$ (circles), $\Delta \phi=0.25$ (upwards triangles), $\Delta \phi=0.3$ (downwards triangles), $\Delta \phi=0.4$ (diamonds), being $\beta\Delta V=10$. The magnitude of the entropic barrier, $\Delta S=0.2,0.4,0.61,0.85,1.1,1,4,1.7,2.2,2.9,3.5,4.6,5.2,6.3,7.6$, is encoded in the color code, the lighter the color of the symbol the larger the value of $\Delta S$.  The dotted line is a guide for the eye highlighting the linear relation between $v$ and $\Gamma$.
\label{fig:simm-simm-no-force-vel}}
\end{figure}

According to Eq.\ref{eq:gamma}, $\Gamma$ depends
on both the number of jumps between the two states and the average
velocity achieved by the motor. Therefore, a further insight into the dynamics
is shown in Fig.~\ref{fig:simm-simm-no-force-vel} that displays the inverse
proportionality  between the average velocity $v$ and $\Gamma$.
 The  dependence displayed in Fig.~\ref{fig:simm-simm-no-force-vel} , in agreement with the simple expectation quantified in  Eq.\ref{eq:gamma}, indicates that the motor 
 hopping  dynamics between the two
states is not significantly affected by the corrugated channel. Therefore the reduction in the energy consumption provided  by cooperative rectification relies on a more efficient transduction of the energy injected in the system into displacement rather than on a reduction of the energy consumed by the motor per unit time. 

\subsubsection*{Symmetric channel and asymmetric ratchet potential}

\begin{figure}
\includegraphics[scale=0.33]{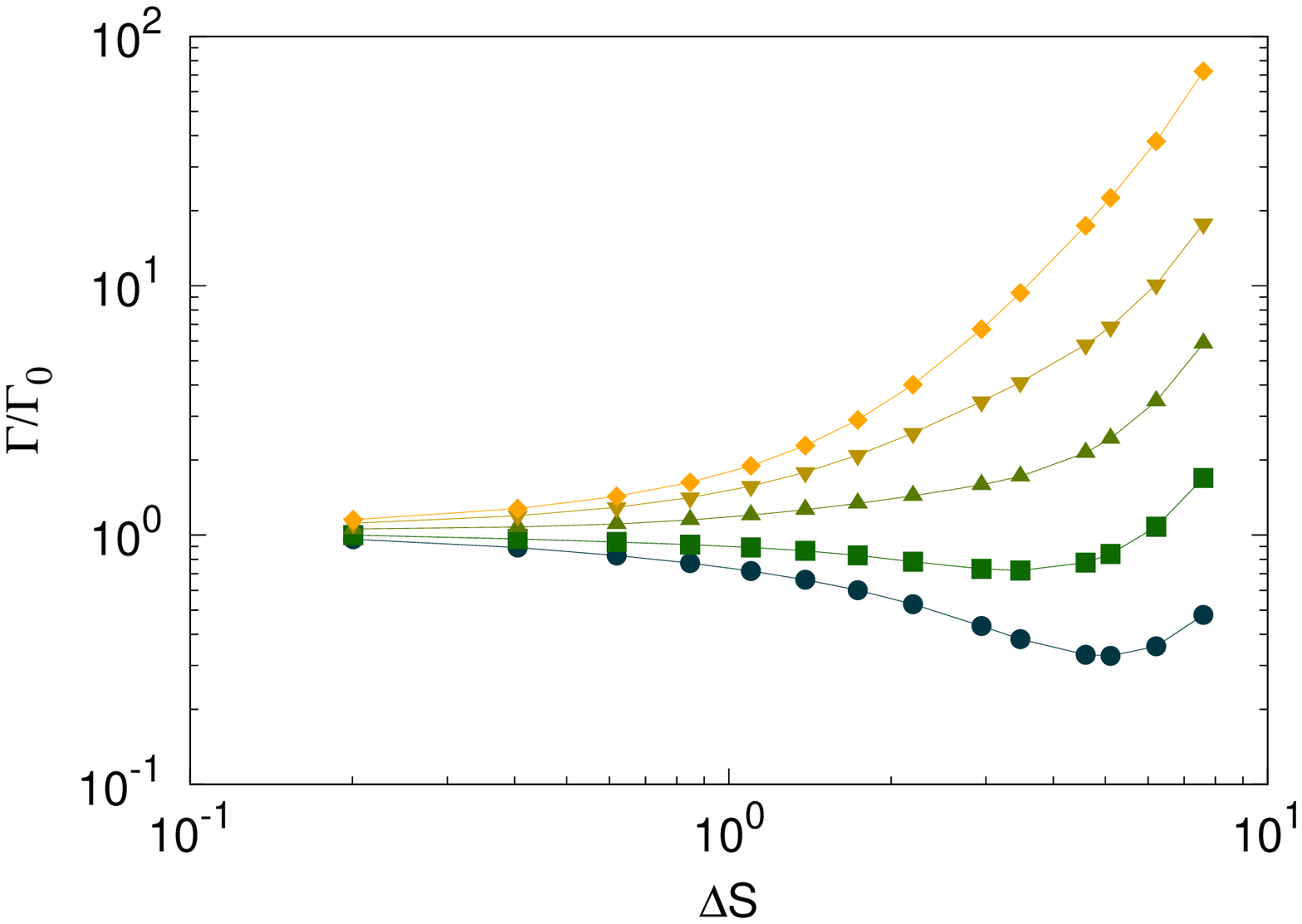} \includegraphics[scale=0.33]{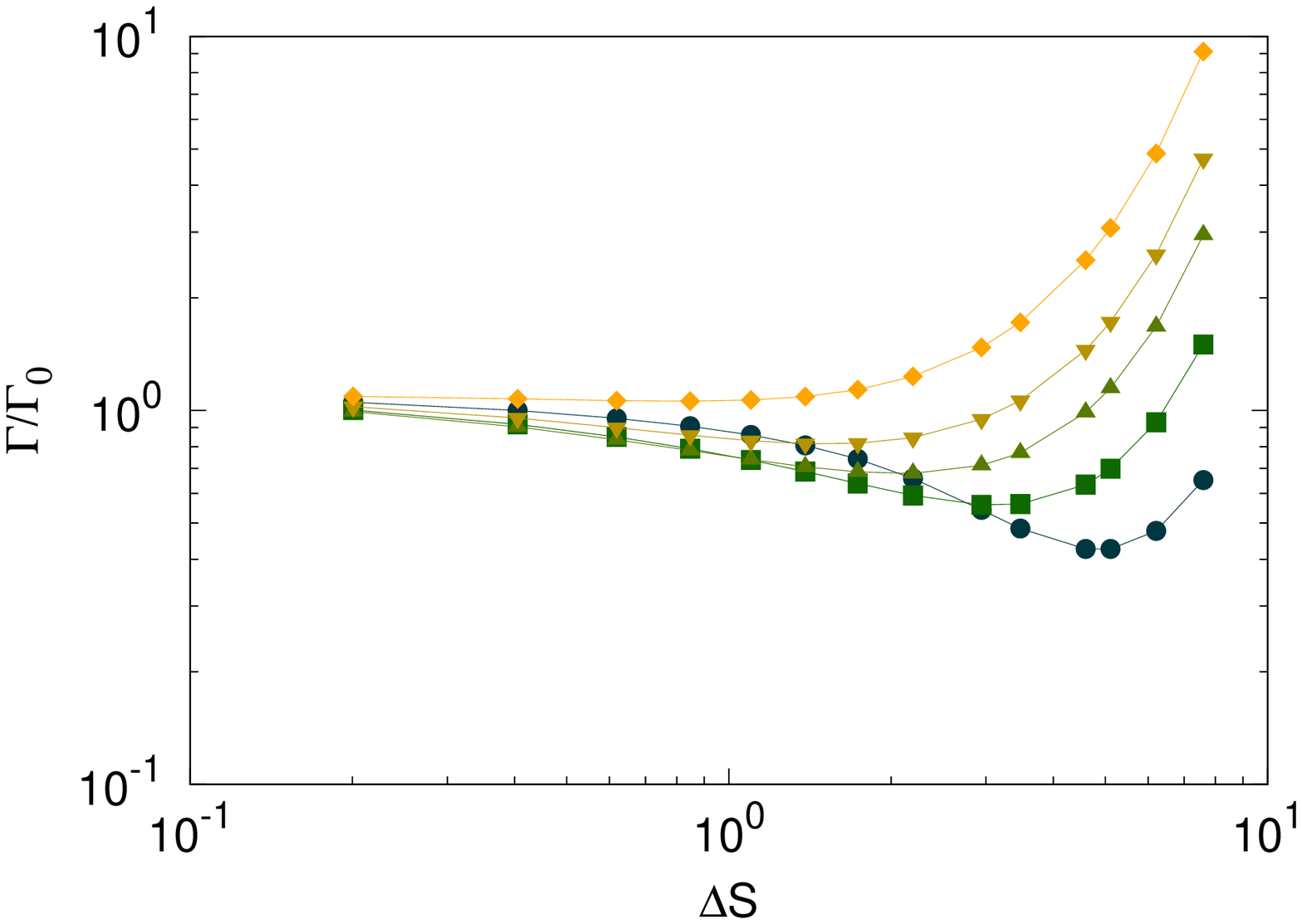}
\caption{$\Gamma$, normalized by $\Gamma(\Delta S=0)=\Gamma_{0}$, as a function
of the entropic barrier, $\Delta S$,  for  processive  (left) and non processive  (right) symmetric molecular motors  moving along a  symmetric channel, for $\beta \Delta V=10$ .  The different curves correspond to different values of the phase shift, $\Delta\phi=0,0.1,0.2,0.3,0.4$
 \label{fig:asimm-simm-no-force}}
\end{figure}

\begin{figure}
\includegraphics[scale=0.33]{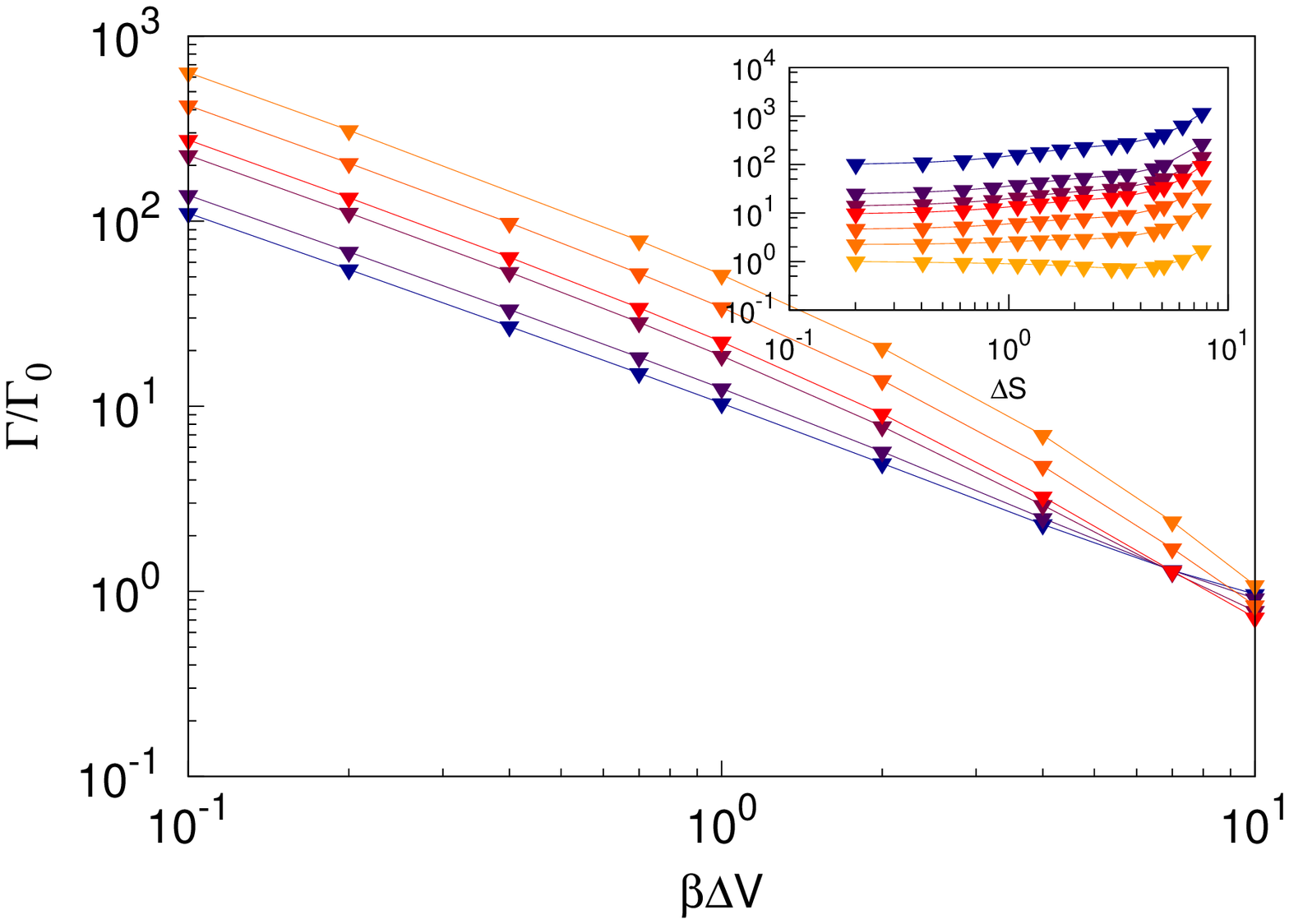} \includegraphics[scale=0.33]{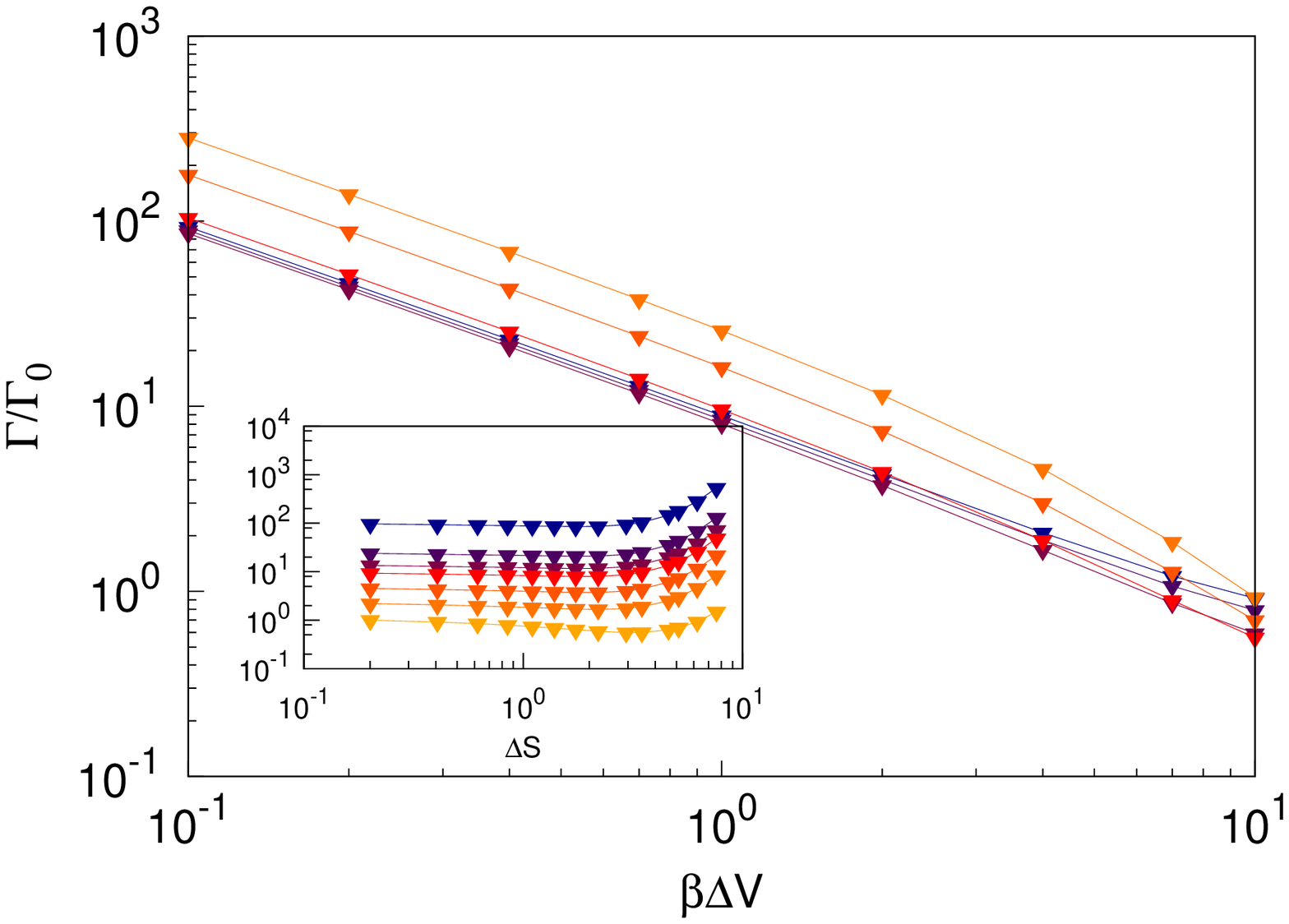}
\caption{$\Gamma/\Gamma_0$ as a function of the  ratchet potential $\Delta V$ for  processive  (left) and non processive  (right) asymmetric molecular motors  moving along a  symmetric channel, for a phase shift $\Delta\phi=0.1$. The different curves correspond to different degrees of confinement, quantified by the magnitude of the  entropic barrier, $\Delta S=0.2,0.85,2.2,3.5,5.1,7.6$; the lighter the color of the curve the larger the value of $\Delta S$.  Inset: $\Gamma/\Gamma_0$ as a function of $\Delta S$ for $\beta \Delta V=0.1,0.4,0.7,1,4,7,10$; the lighter the color of the curve the larger the value of  $\Delta V$. \label{fig:asimm-simm-no-force-velpot}}
\end{figure}

Molecular motors will now displace  even in the absence of a spatially-varying channel.
 Therefore, we can define $\Gamma_0$ as the value of $\Gamma$ obtained for a flat channel, i.e. for $\Delta S=0$. In this way we can characterize the deviations in $\Gamma$ due to the geometrical confinement. As it has already been discussed~\cite{Malgaretti2012,Malgaretti2013}, the presence of geometrical constraints strongly affect the net motion
of molecular motors even when motors can rectify by themselves. Fig.~\ref{fig:asimm-simm-no-force} shows the
dependence of $\Gamma$ on the entropic barrier $\Delta S$. For both
processive and non processive motors, $\Gamma$ shows a non monotonous
behavior identifying an optimal entropic barrier for which $\Gamma$ attains
a minimum. Therefore the presence of the entropic barrier not only
tunes motor currents~\cite{Malgaretti2012,Malgaretti2013} but also allows to reduce the energy
consumed by motors in order to perform a single step. 

\begin{figure}
\includegraphics[scale=0.33]{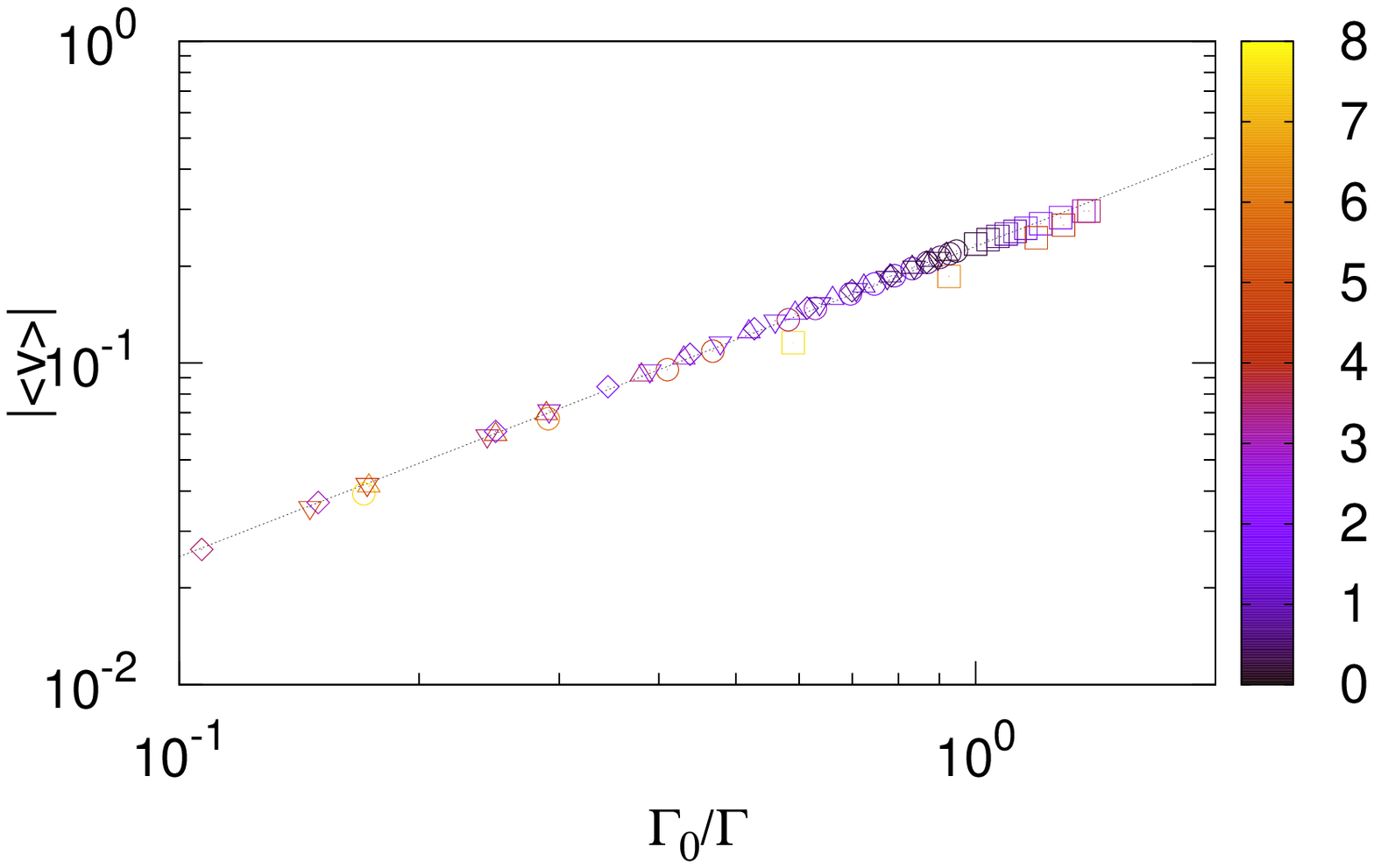} \includegraphics[scale=0.33]{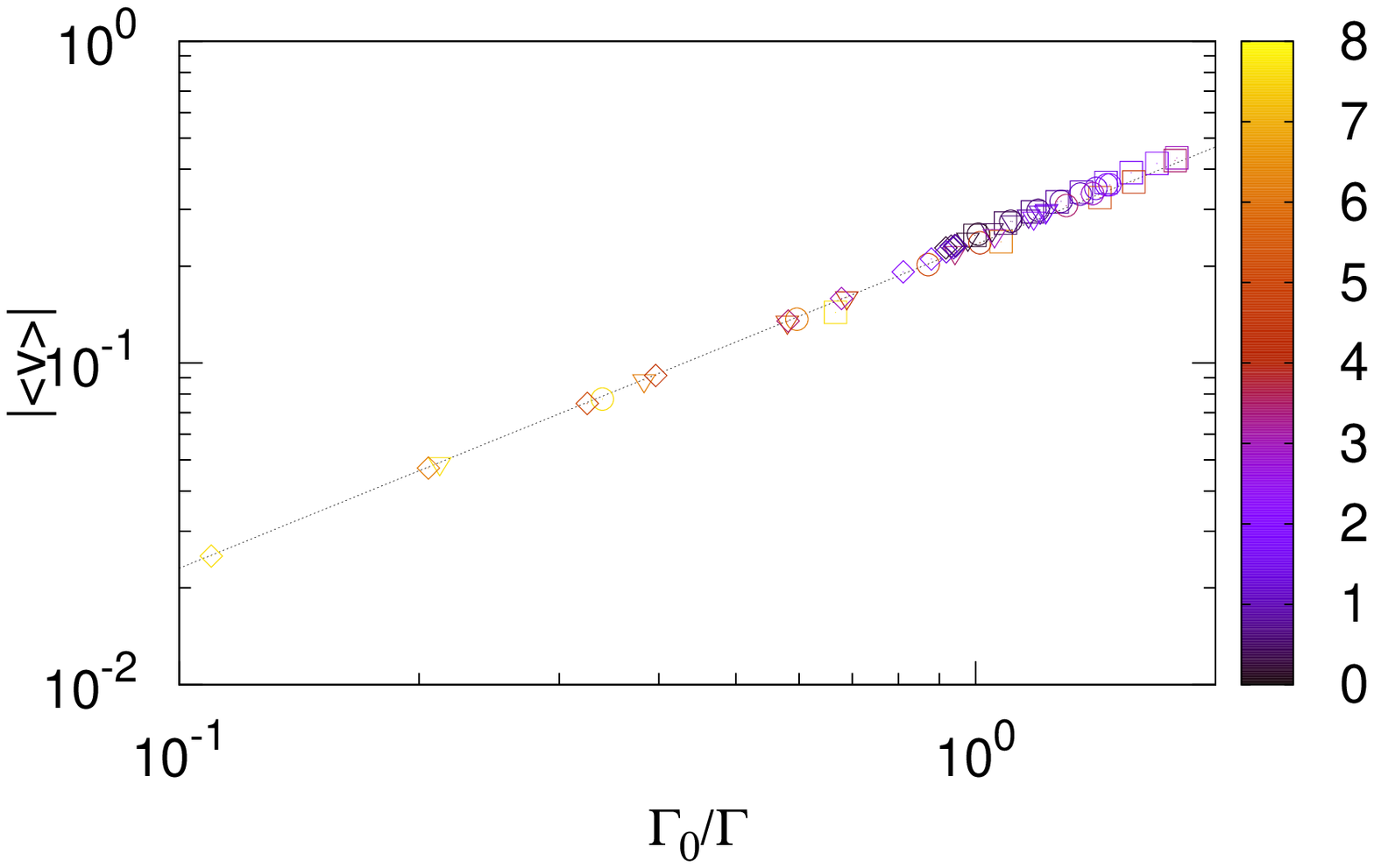}
\caption{Mean molecular motor velocity, $v$, as a function of $\Gamma/\Gamma_0$ of  processive (left) and non processive (right) asymmetric molecular  motors displacing in a spatially-varying, symmetric channel. The different symbols correspond to different phase shifts, quantified by $\Delta \phi=0.1$ (squares), $\Delta \phi=0.2$ (circles), $\Delta \phi=0.25$ (upwards triangles), $\Delta \phi=0.3$ (downwards triangles), $\Delta \phi=0.4$ (diamonds), being $\beta\Delta V=10$. The magnitude of the entropic barrier $\Delta S=0.2,0.4,0.61,0.85,1.1,1,4,1.7,2.2,2.9,3.5,4.6,5.2,6.3,7.6$, is encoded in the color of the symbols, the lighter the color of the curve the larger the value of $\Delta S$. The dotted line is a guide for the eye highlighting the linear relation between $v$ and $\Gamma$.\label{fig:asimm-simm-no-force-vel}}
\end{figure}

As shown in Fig.~\ref{fig:asimm-simm-no-force-velpot} the behavior of $\Gamma/\Gamma_0$ as a function of $\Delta V$ is similar to that observed for a symmetric ratchet potential, see Fig.~\ref{fig:simm-simm-no-force-velVspot}. However, in the present case the value of $\Delta S$ for which the minimum of $\Gamma$  is more sensitive to  $\Delta V$. For smaller values of $\Delta V$ the minimum of $\Gamma$ is obtained by minimizing $\Delta S$, while for larger values of $\Delta V$ the minimum is obtained for non vanishing $\Delta S$. As in the case of a symmetric ratchet potential, we find an inverse proportionality
relation between $\Gamma$ and $v$, shown in Fig.~\ref{fig:asimm-simm-no-force-vel}, indicating that, analogously to the case of symmetric ratchet potential, the hopping dynamics of the molecular motor between its two internal states  is not strongly affected by the confinement imposed by the corrugated channel. 

\subsection*{Motor performance in the presence of an external force}

When motors are pulling against an external force it is possible to define a thermodynamic efficiency, as discussed to introduce Eq.~\ref{efficiency}. 
We can also analyze the performance of a symmetric motor in a symmetric channel or when the   ratchet potential is asymmetric. In this latter case the motor already displace when moving along a flat channel. Hence, in this second scenario it is possible to define a ``bulk'' efficiency, $\eta_{0}$, which we use as a reference to characterize the role of the geometrical constraint on motor's efficiency. 

The external force decreases the intrinsic motor velocity and it can eventually invert its motion. Obviously, at the stall force the molecular motor efficiency vanishes and the motor efficiency at larger values of the force is not well defined. Hence, we will not analyze motors efficiency beyond the stall force.

\begin{figure}
\includegraphics[scale=0.33]{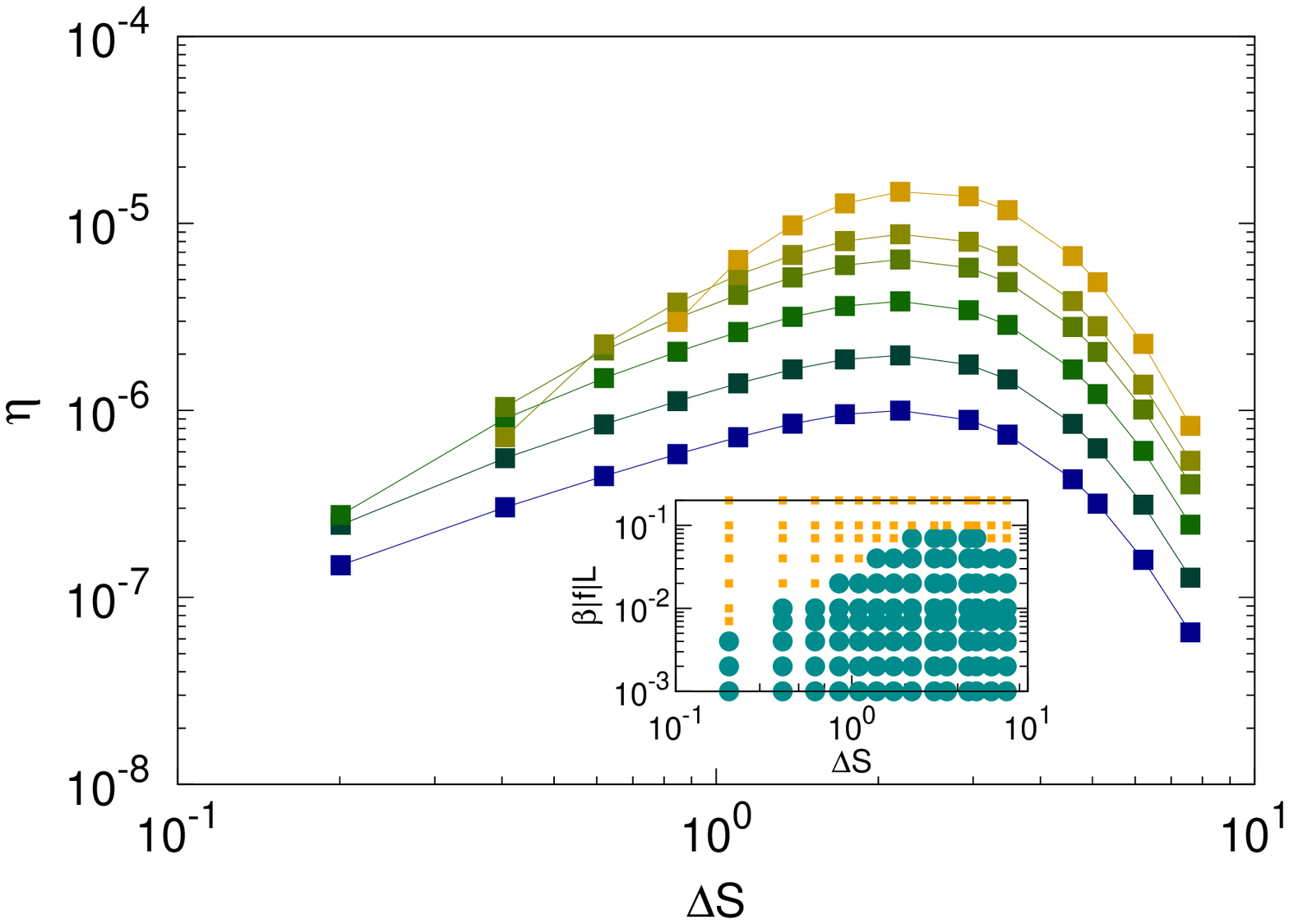} \includegraphics[scale=0.33]{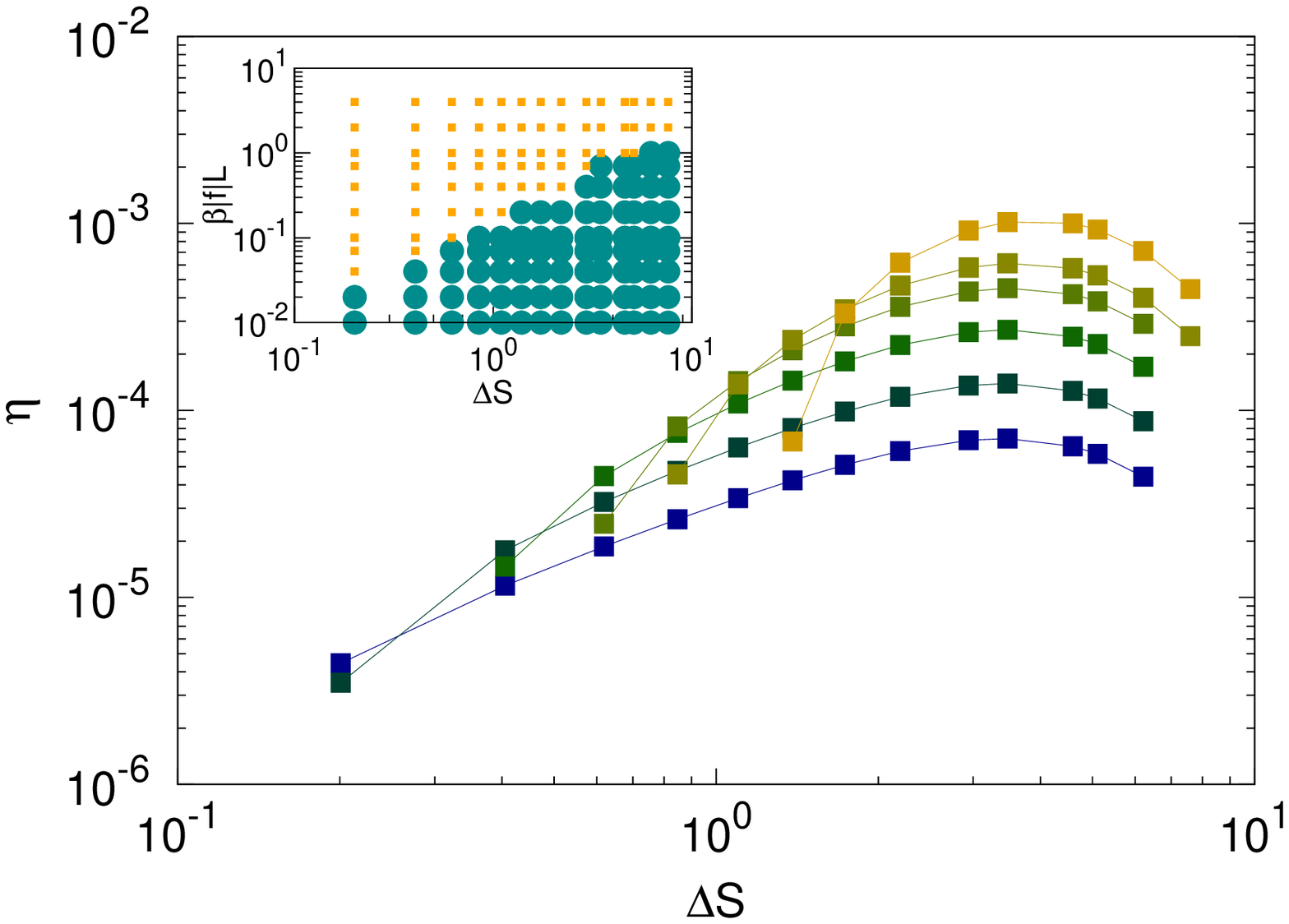}
\caption{Molecular motor efficiency,$\eta$, as a function of the entropic barrier $\Delta S$ for processive  (left) and non processive  (right) symmetric molecular motors moving along a spatially-varying, symmetric channel, for a phase shift  $\Delta\phi=0.2$ and $\beta\Delta V=10$. The different curves correspond to different values of the applied external force,  $\beta|f|L=0.001,0.002,0.004,0.007,0.01,0.02$; the lighter the color of the curve the larger the magnitude of $f$.  Insets: stepping state of the motor. Bigger circles represent the sets of parameters $\Delta S$ and $f$ for which the motor can step against the force, smaller squares the set of values for which the motor cannot step against the force}
\label{fig:simm-simm-force}
\end{figure}

\subsubsection*{Symmetric channel and ratchet potential}

Since molecular motor rectification emerges from the interplay between molecular  motor and channel corrugation, symmetric molecular motors cannot  displace against the applied force for a flat channel. As the corrugation increases there exists a finite stall force that initially increases with channel corrugation. Fig.~\ref{fig:simm-simm-force} and its insets show that both processive and non-processive motors have a maximum stall force for a finite, optimal channel corrugation. For the same reason, increasing the magnitude of the applied force decreases the range of channel corrugations where the motor displaces against the applied force, as shown in Fig.~\ref{fig:simm-simm-force}. Moreover, the insets Fig.~\ref{fig:simm-simm-force} also show that there exists a maximal stall force beyond which the motor cannot displace  against the applied force, independently of the channel corrugation.
Comparing the insets in Fig.~\ref{fig:simm-simm-force} with their corresponding main panels, we find that the stall force is maximized at maximum efficiency for both processive and non processive motors.

 Fig.~\ref{fig:simm-simm-force} shows the dependence of the efficiency, $\eta$, as a function of  the entropic barrier $\Delta S$. For both processive and non processive motors we observe an optimum of the efficiency for non vanishing values of $\Delta S$. Such a behavior derives from the nontrivial dependence of cooperative rectification shown in Fig.~\ref{fig:simm-simm-no-force} where we can identify an optimal value of the entropic barrier, $\Delta S_{opt}$ that minimizes $\Gamma$. Such a behavior is retained leading to a maximum efficiency for $\Delta S\simeq\Delta S_{opt}$. 
Therefore the reduction of the energy cost observed in the absence of external forces, see Fig.~\ref{fig:simm-simm-no-force}, leads to an increase of the efficiency when motors are pulling against applied forces.

\begin{figure}
\includegraphics[scale=0.33]{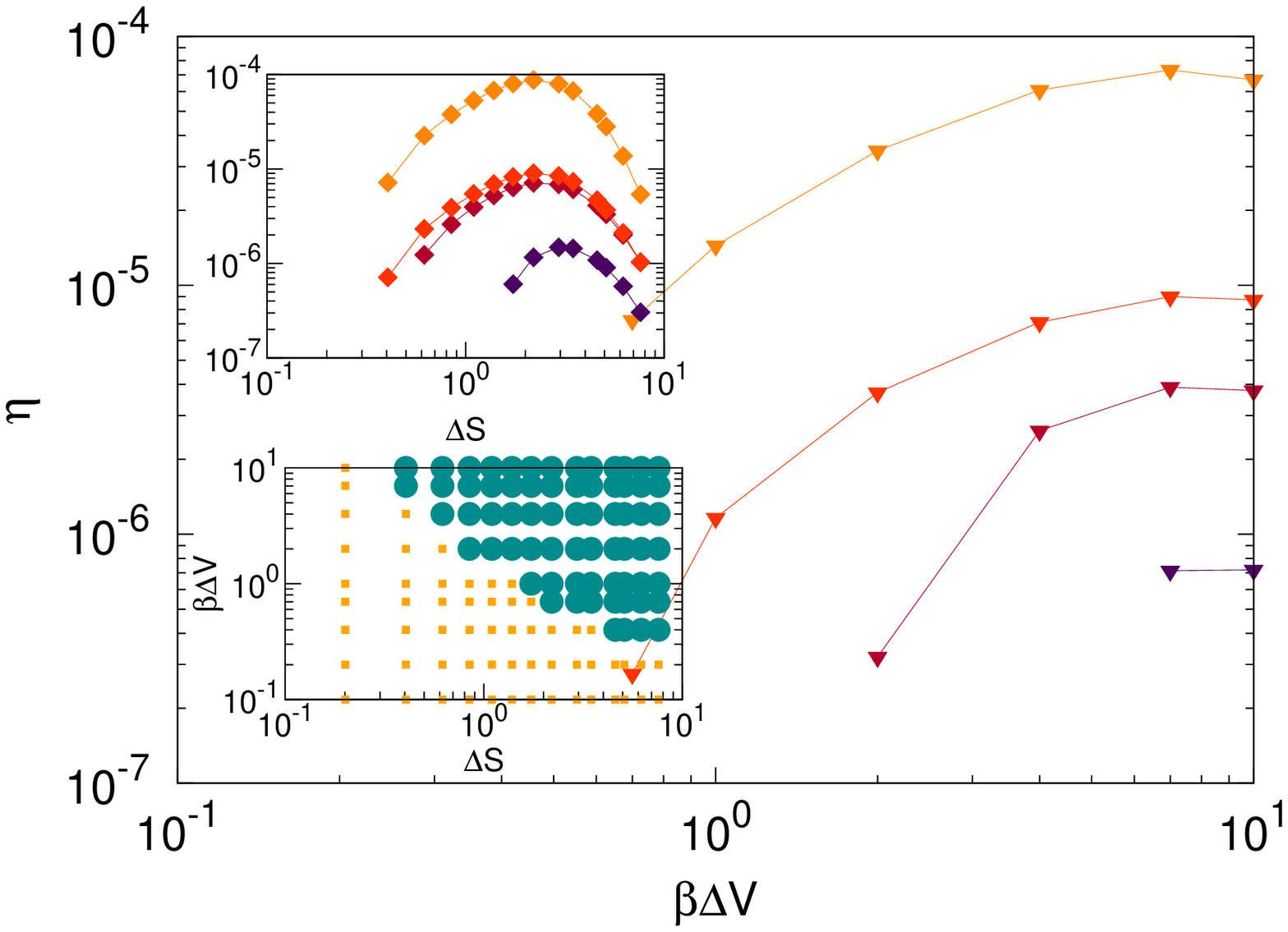} \includegraphics[scale=0.33]{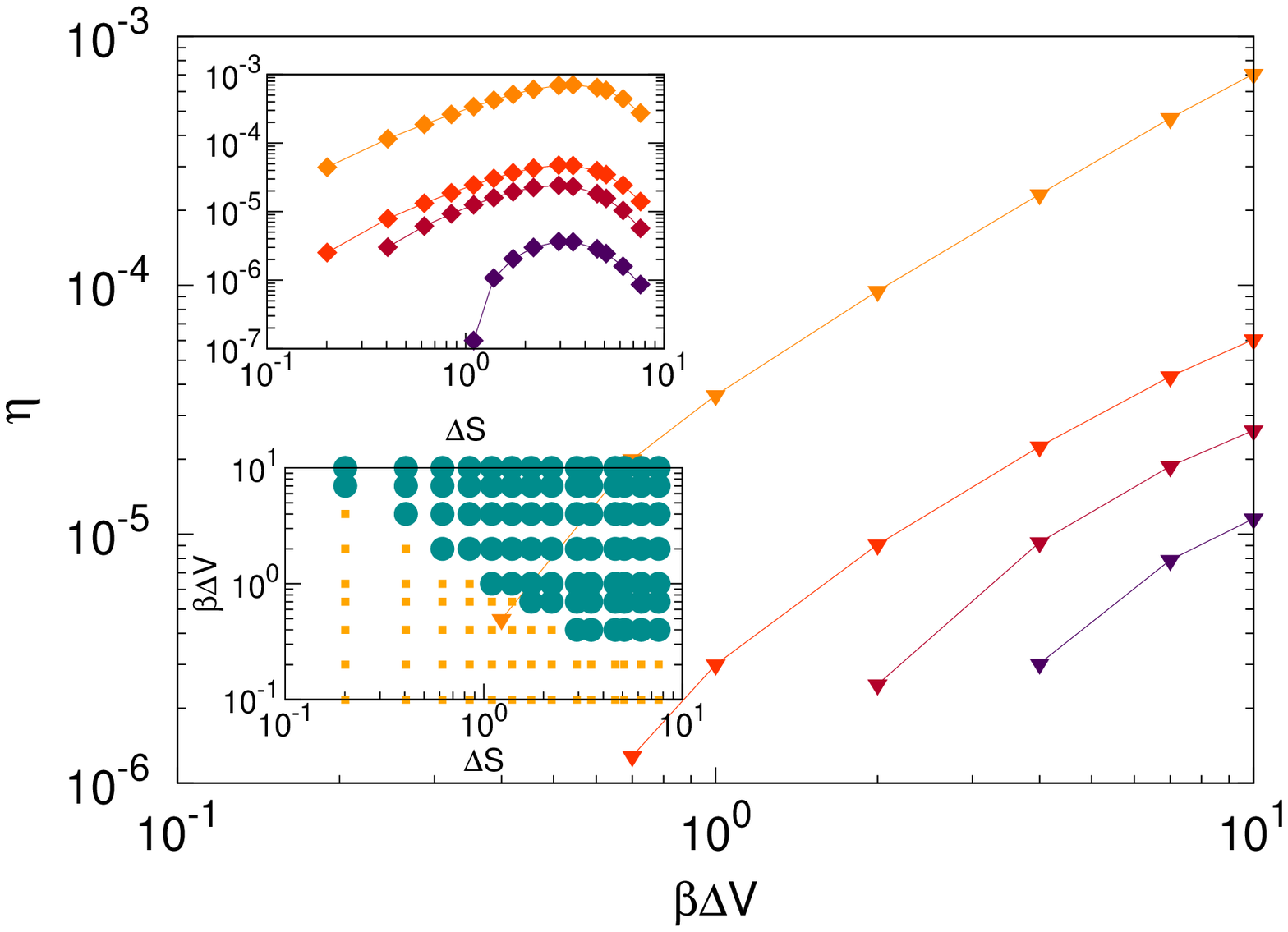}
\caption{Molecular motor efficiency,$\eta$, as a function of the ratchet potential depth $\Delta V$ for processive  (left) and non processive  (right) symmetric molecular motors moving along a spatially-varying, symmetric channel, for a phase shift  $\Delta\phi/(2 \pi)=0.2$. The different curves correspond to different values of the entropic barrier $\Delta S=0.4,0.85,2.2,3.5$; the lighter the color of the curve the larger the magnitude of $\Delta S$. Upper insets: efficiency as a function of $\Delta S$ for $\beta \Delta V=1,4,7,10$ for $\Delta \phi=0.2$. Lower insets: stepping state of the motor. Bigger circles represent the sets of parameters $\Delta S$ and $f$ for which the motor can step against the force, smaller squares the set of values for which the motor cannot step against the force. 
\label{fig:simm-simm-force_velVSpot}}
\end{figure}


Fig.~\ref{fig:simm-simm-force_velVSpot} and its top insets show the dependence of the efficiency upon variations of the ratchet potential depth, $\Delta V$ for different values of $\Delta S$. As compared to the case of absence of external force, see Fig.~\ref{fig:simm-simm-no-force-velVspot}, here we have found a more sensitive dependence of the efficiency on $\Delta V$. 

When the motor step against the force, we have found that the value of $\Delta V$ maximizing efficiency is quite sensitive to the value of the entropic barrier $\Delta S$ as can be see in the main figure and in the upper insets. Moreover processive and non processive motors show a diverse relation between the efficiency and $\Delta V$.  In fact, for processive motors we observe that $\eta$ saturates when increasing $\Delta V$ while for non processive motors we observe a monotonous growth of  $\eta$  with $\Delta V$  for the range of  values of $\Delta V$ explored, as  shown in Fig.~\ref{fig:simm-simm-force_velVSpot}. 
 
The dependence of the efficiency on the force is more involved. As shown in Fig.~\ref{fig:simm-simm-force}, when $\Delta S$ maximizes the efficiency, i.e. for $\Delta S=\Delta S_{opt}$, we find that $\eta$ increases with the external force. At larger values of the external force, when it approaches the stall force, the efficiency will decrease and will eventually vanish for the corresponding corrugation, $\Delta S$. Fig.~\ref{fig:simm-simm-force} shows that for   $\Delta S< \Delta S_{opt}$  the maximum efficiency is not always attained for the largest of the range of applied external forces sampled, showing already the existence of a characteristic external force for which  the optimal efficiency is achieved; this optimal force changes with the channel corrugation, $\Delta S$.

\begin{figure}
\includegraphics[scale=0.33]{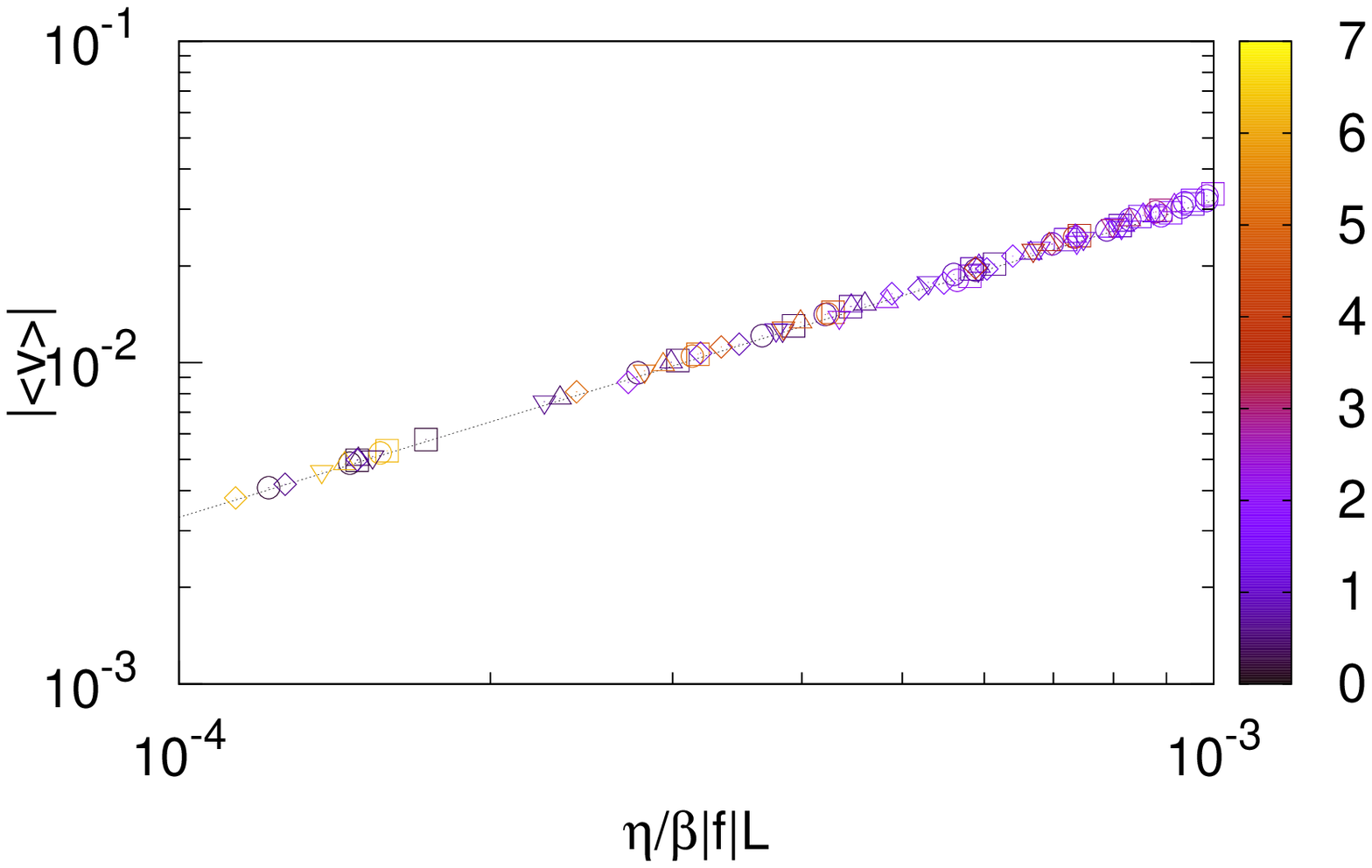} \includegraphics[scale=0.33]{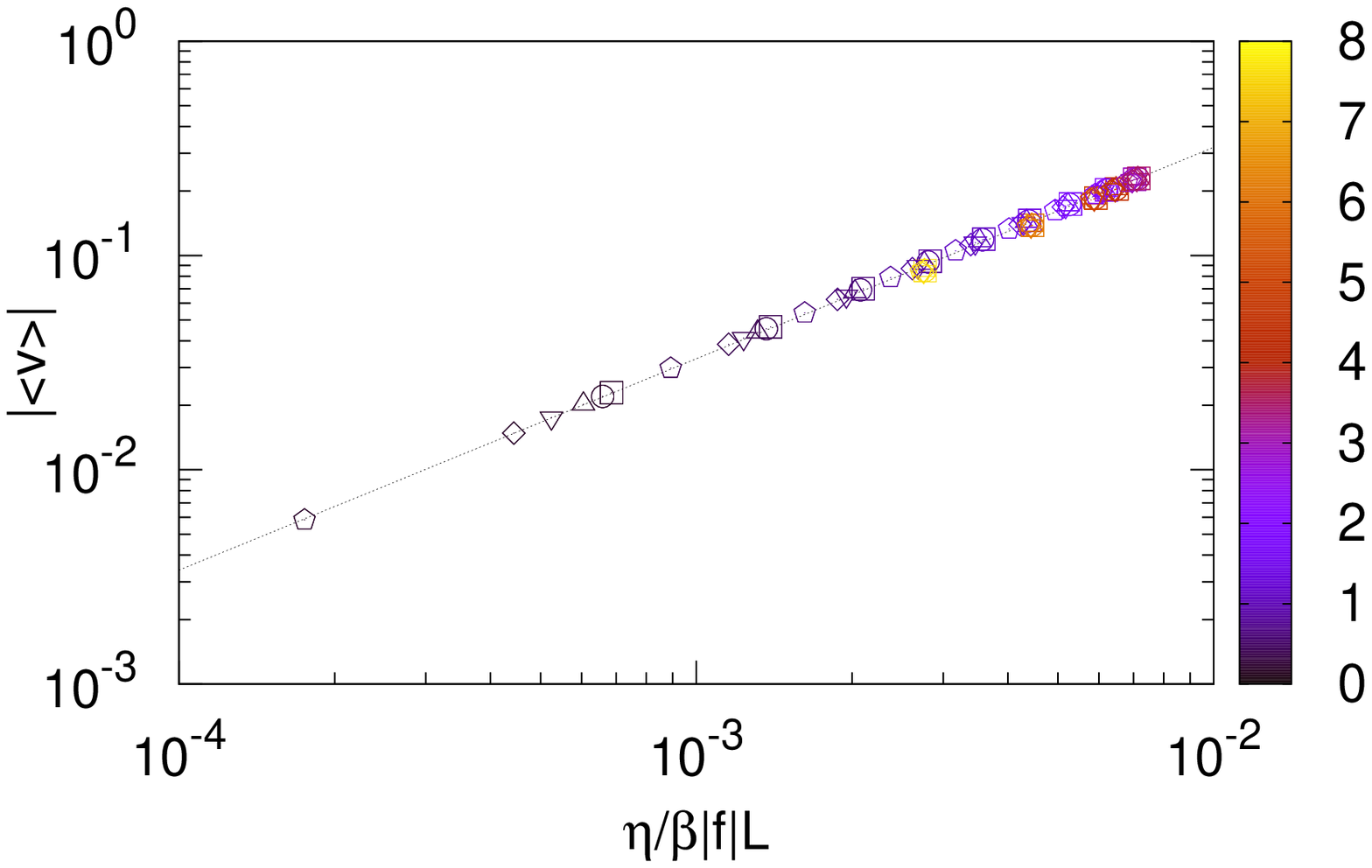}
\caption{ Mean velocity, $v$, as a function of the relative efficiency, $\eta/\eta_0$, of a processive (left) and non processive (right) motor for $\delta \phi=0.2$ and $\beta \Delta V=10$. Different values of the external force $\beta |f| L=0.001,0.002,0.004,0.007,0.01,0.02,0.04$ are encoded by different simbols (squares, circles, up triangles, down triangles, diamonds, pentagons, crosses respectively) while the different values of $\Delta S$ are  color coded, where lighter colors stand for larger values of $\beta |f| LS$.
\label{fig:simm-simm-force-vel}}
\end{figure}

However, the relation between $\eta$ and the average velocity, $v$, is linear, as shown in Fig.~\ref{fig:simm-simm-force-vel}. Moreover, similarly to what we have seen in the previous cases\footnote{We remind that  $\eta\propto\Gamma^{-1}$}, all the data collapse on a straight line ensuring a direct proportionality between $\eta$ and $v$. 
Coherently with the previous cases, we conclude that the hopping dynamics is only slightly affected by the geometrical confinement and the main dependence of the efficiency on $\Delta S$ comes from the velocity.

\begin{figure}
\includegraphics[scale=0.33]{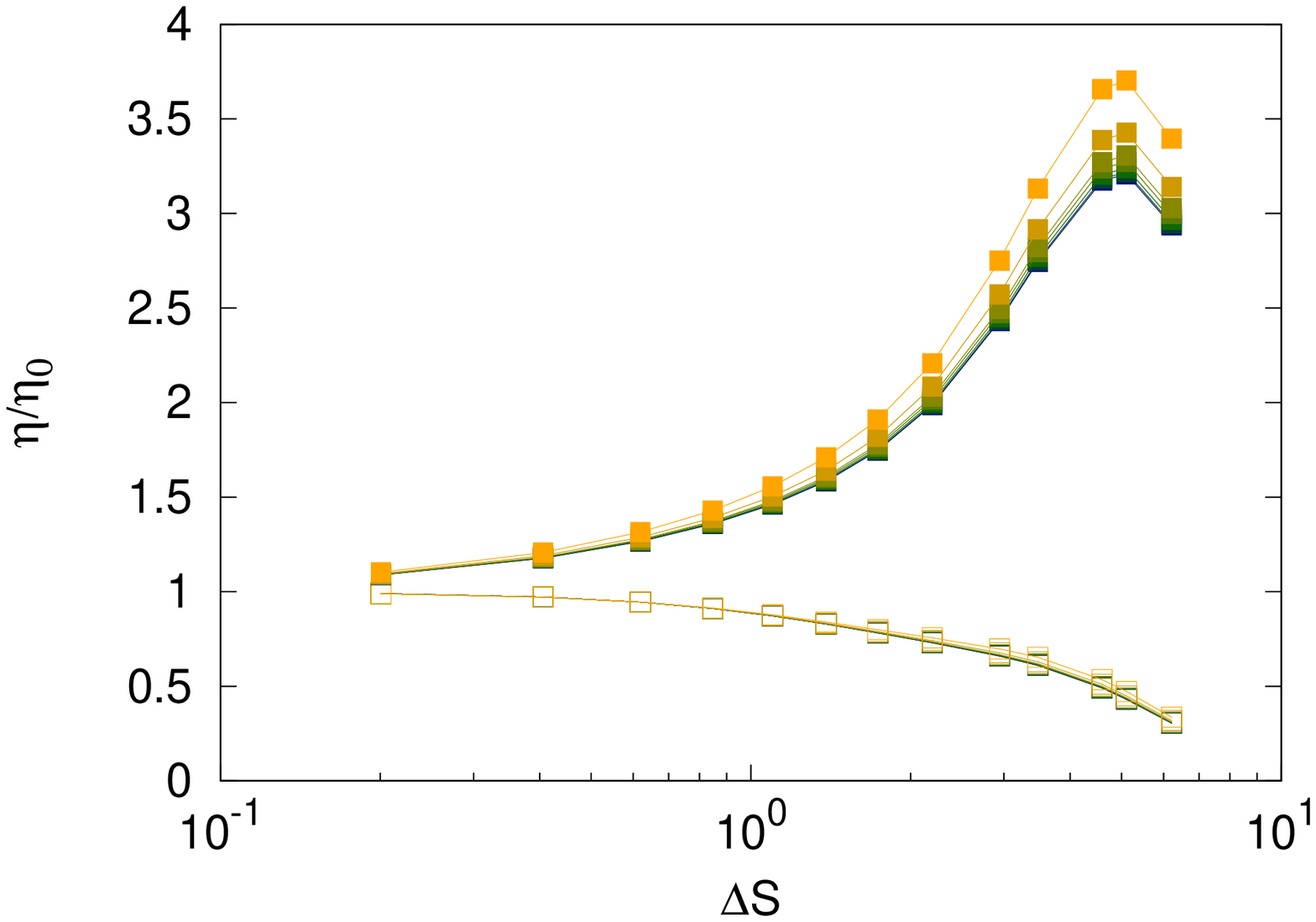} \includegraphics[scale=0.33]{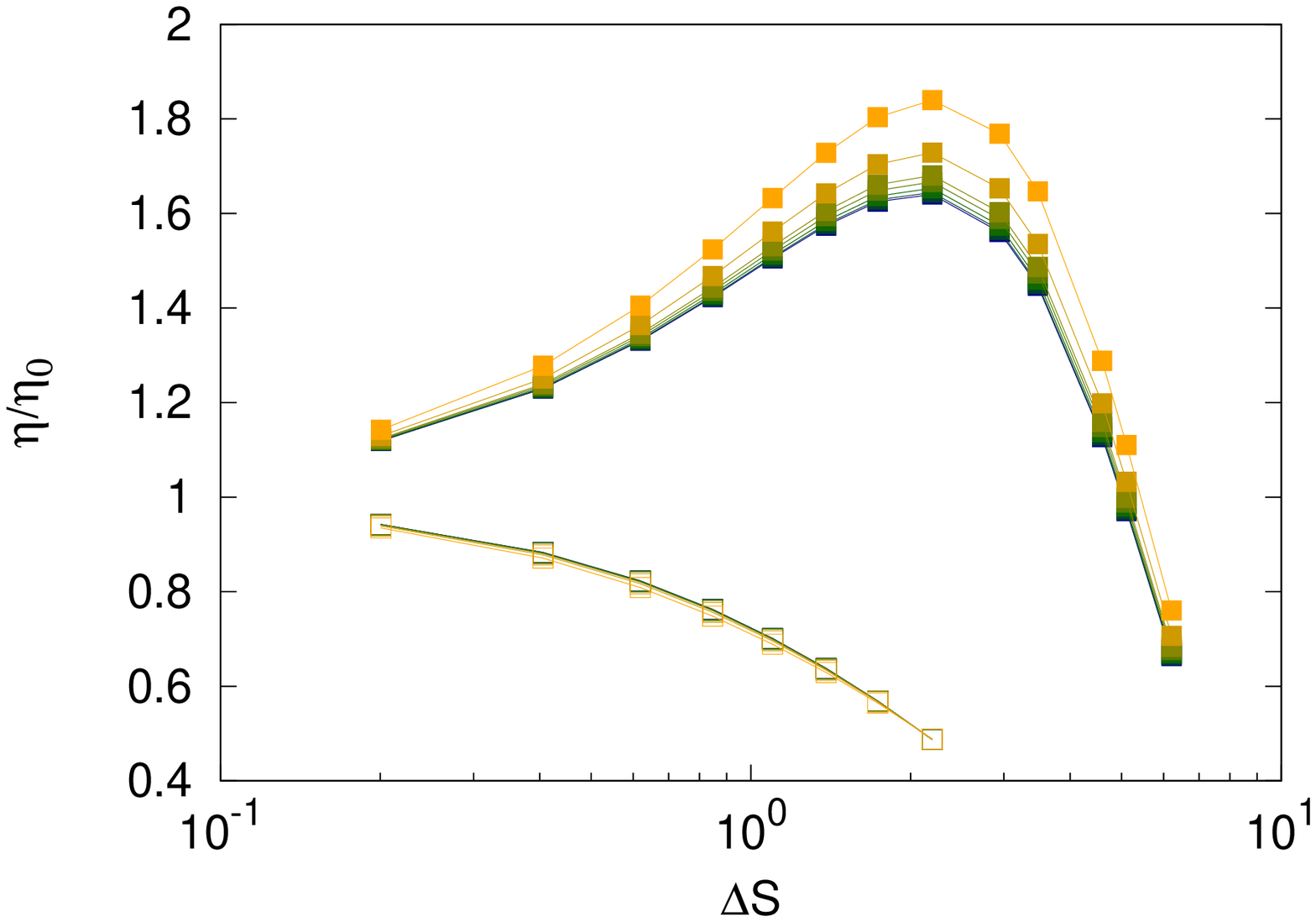}
\caption{$\eta$, normalized by the efficiency $\eta_{0}$ of a motor in a flat
channel, as a function of the entropic barrier $\Delta S$ for different
values of the force, $f$, where lighter lines stands for larger values
of $f$, $\beta|f|L=0.001,0.002,0.004,0.007,0.01,0.02$, for a processive
motor (left) and non processive motor (right) in the case of an asymmetric
ratchet potential and symmetric channel shape. The phase shift $\Delta\phi$
is $\Delta\phi=0$ ($\Delta\phi=0.2$) for filled (empty) points in the case of
processive motor and $\Delta\phi=0.2$ ($\Delta\phi=0.5$) for filled (empty)
points in the case of non processive motors.} 
\label{asimm-simm-force}
\end{figure}

\begin{figure}
\includegraphics[scale=0.33]{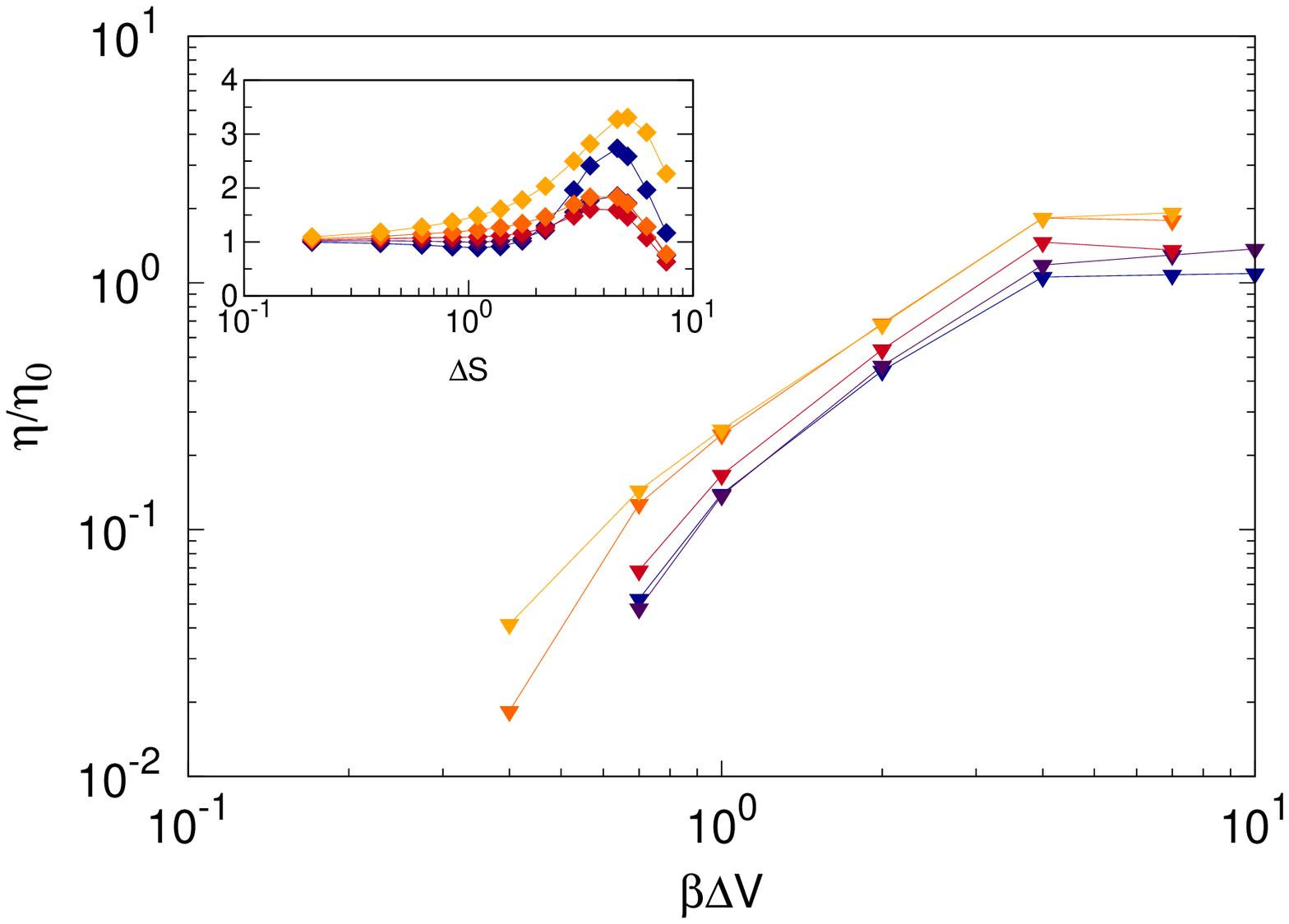} \includegraphics[scale=0.33]{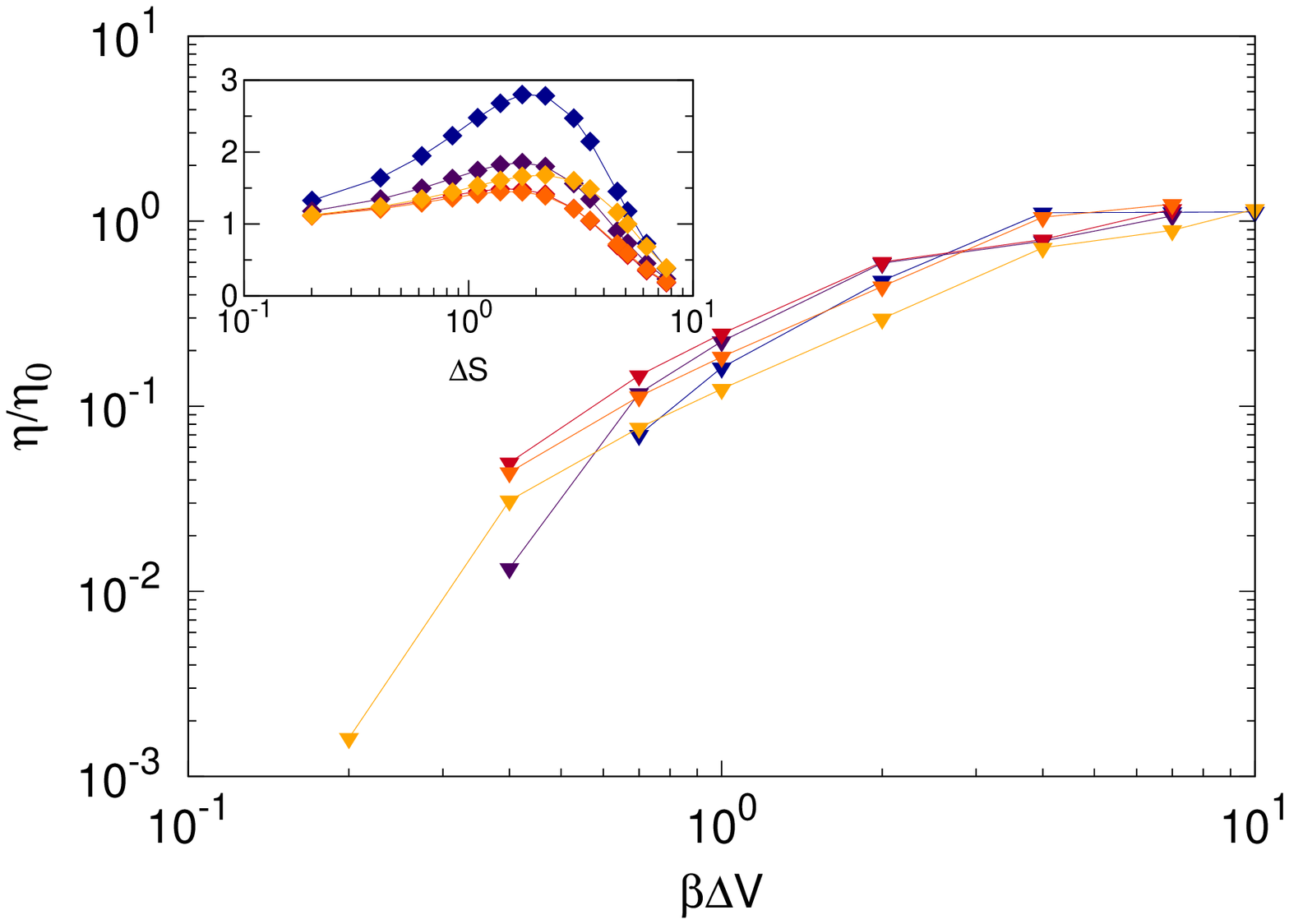}
\caption{$\eta/\eta_0$ as a function of the entropic barrier $\Delta S$ for different
values of the ratchet potential $\Delta V=0.1,0.2,0.4,0.7,1,2,4,7,10$, where lighter lines stands for larger values
of $\Delta V$ for a processive motor (left) and non processive motor (right) in the
case of a symmetric ratchet potential and symmetric channel shape, being $\Delta\phi=0.$ for processive motors and $\Delta\phi=0.2$ for non processive motors.
\label{fig:asimm-simm-force_velVSpot}}
\end{figure}
\subsubsection*{Symmetric channel and asymmetric ratchet potential}

 When the ratchet potential is asymmetric, motors can step in a flat channel,  $\Delta S=0$. Therefore, if the external force is smaller than the stall force, we can define a reference efficiency, $\eta_0$ defined as the efficiency obtained for $\Delta S=0$. 
Fig.~\ref{asimm-simm-force} shows the dependence of the normalized efficiency upon variation of the entropic barrier $\Delta S$. Both processive and non processive motors show a confinement-induced efficiency enhancement and an optimal value of $\Delta S$. Such a behavior is strongly affected by the phase shift as well as by the processive nature of the motors. In fact, while for $\Delta\phi=0$ ($\Delta\phi=0.2$) for processive (non processive) motors we observe an enhancement in the efficiency upon increasing $\Delta S$, for $\Delta\phi=0.2$ ($\Delta\phi=0.5$) we observe a reduction in the efficiency for increasing values of $\Delta S$.

The dependence of $\eta/\eta_0$ on $\Delta V$ is shown in Fig.~\ref{fig:asimm-simm-force_velVSpot}. Differently from the case of symmetric ratchet potential, in the present case both processive and non processive motors exhibit a monotonous increase of the efficiency with $\Delta V$ till $\beta\Delta V\simeq10$ where $\eta/\eta_0$ shows a plateau. As in the previous case, see Fig.~\ref{fig:simm-simm-force_velVSpot}, the efficiency increases for larger $\Delta S$. The insets of Fig.~\ref{fig:asimm-simm-force_velVSpot} show the dependence of the efficiency as a function of $\Delta S$ for different values of $\Delta V$. As shown in the figure, for vanishing small corrugation, $\Delta S \rightarrow 0$ we find $\eta/\eta_0\rightarrow 1$. However, by increasing $\Delta S$ the efficiency shows a non monotonous behavior and it reaches a maximum for a finite value of $\Delta S$. 

\begin{figure}
\includegraphics[scale=0.33]{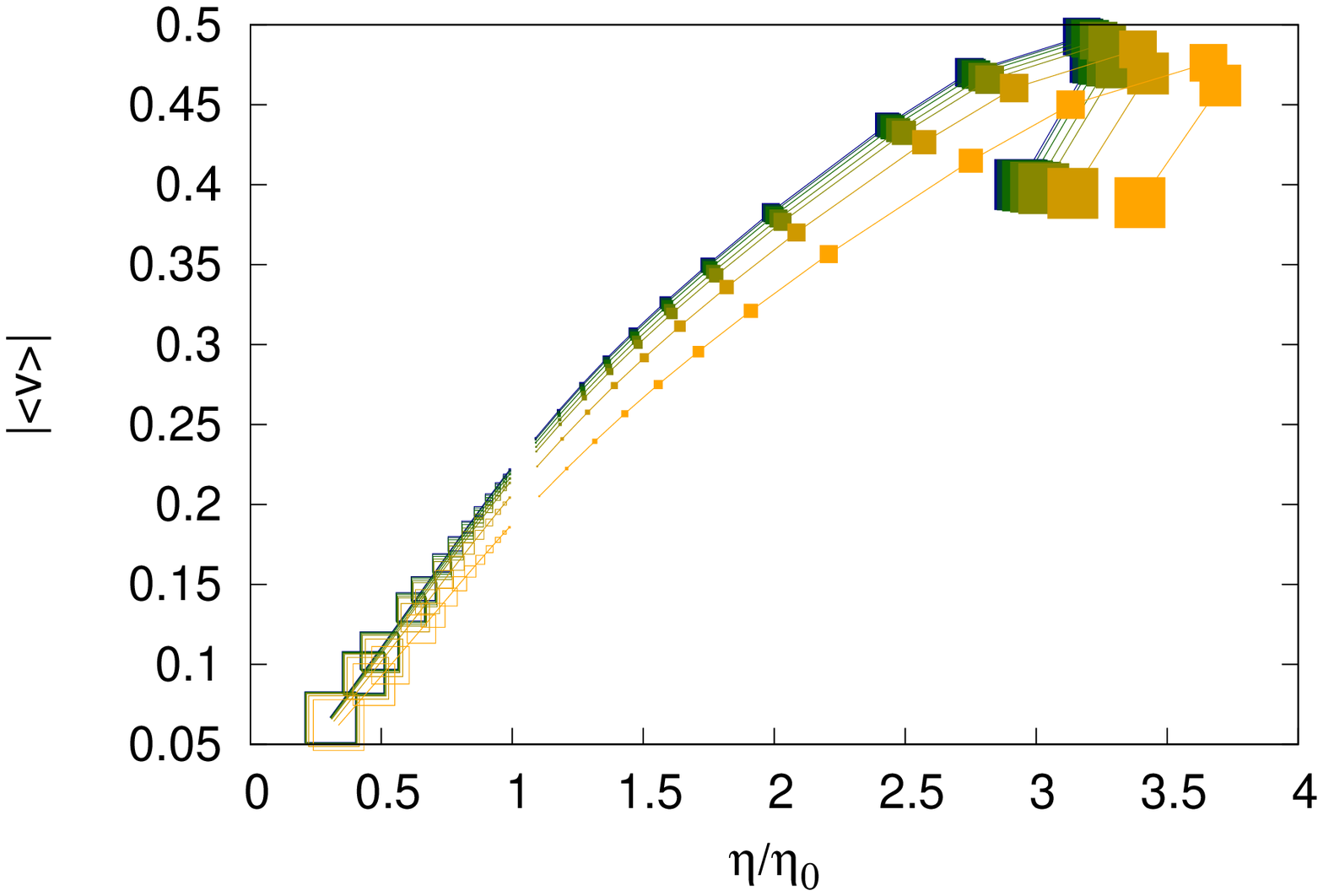} \includegraphics[scale=0.33]{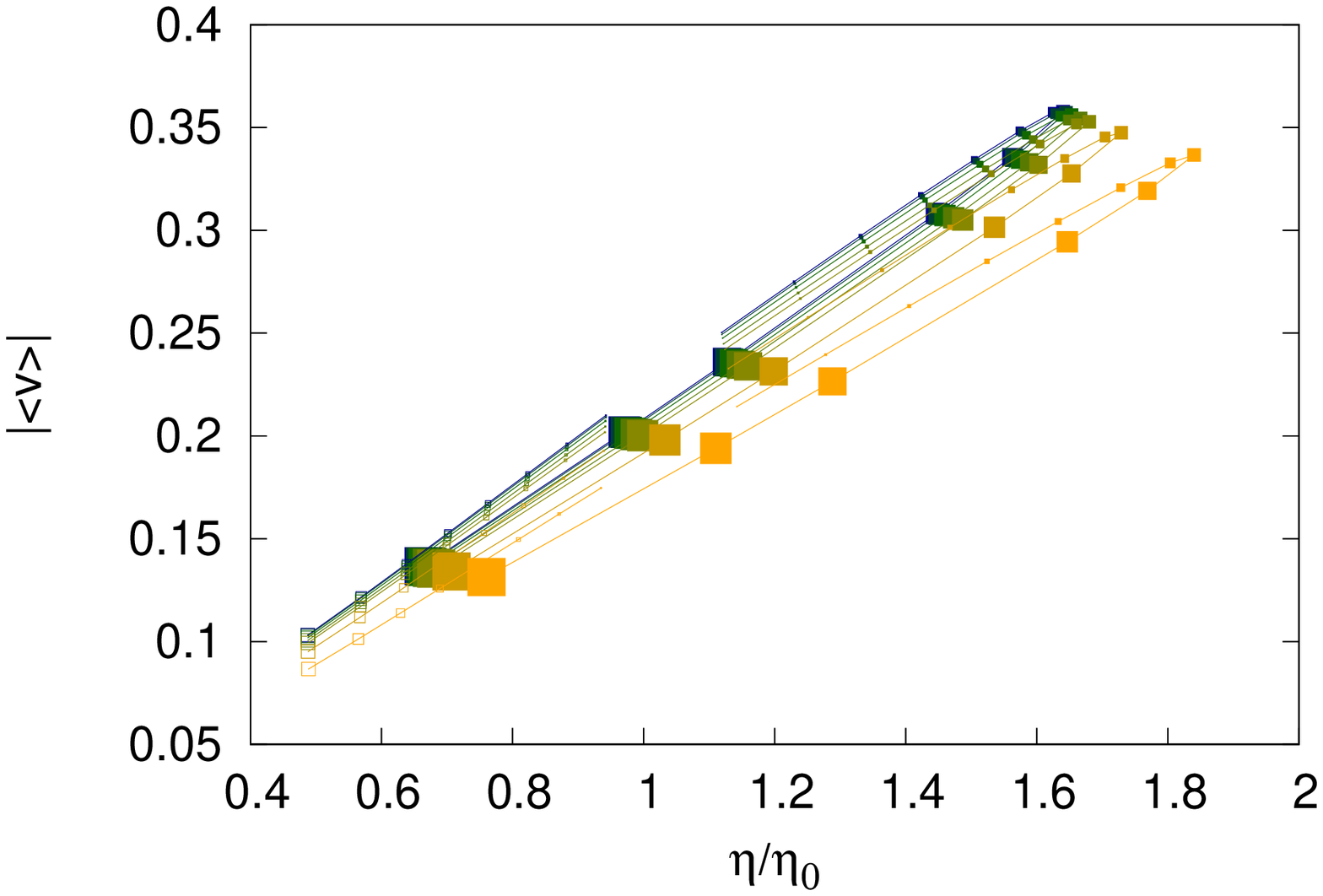}
\caption{Mean velocity, $v$, as a function of the relative efficiency, $\eta/\eta_0$, of a processive (left) and non processive (right) motor for $\beta |f| L=0.001,0.002,0.004,0.007,0.01,0.02,0.04$ (color code) where lighter colors stand for larger values of the force and for $\Delta S =0.2,0.4,0.61,0.85,1.1,1,4,1.7,2.2,2.9,3.5,4.6,5.2,6.3,7.6$ where bigger symbols stand for larger values of $\Delta S$.
\label{fig:asimm-simm-force-vel}}
\end{figure}

 Finally the relation between the efficiency and the net velocity is shown in Fig.~\ref{fig:asimm-simm-force-vel}. Surprisingly in the present case we find, for both processive and non processive motors, a non linear relation between $\eta/\eta_0$ and $v$ as opposed to all other cases analyzed previously. Fig.~\ref{fig:asimm-simm-force-vel} shows that the maximum speed is obtained at maximum efficiency. However, when either efficiency or velocity are not maximized, we find that for a given efficiency motors can achieve two different speeds and viceversa. The value of entropic barriers that maximizes the efficiency, $\Delta S_{opt}$, identifies the threshold between two scenarios and for $\Delta S<\Delta S_{opt}$ motors displace less efficiently as compared to motors moving at the same speed in a channel characterized by a larger value of  $\Delta S$.

\section*{Conclusions}

We have shown that the presence of geometrical constraints strongly affect the performance of molecular motors. In particular the efficiency, or the energy spent per unit step, shows a strong dependence on the environmental properties, encoded in the entropic barrier. 

When motors are not pulling against an externally applied force there is no clear definition of efficiency since motors are not producing any mechanical work. However, it is possible to capture their performance by measuring the energy cost, $\Gamma$, needed to perform a single step forward. We have shown that the cooperative rectification that arises from  the interplay between  internal, out of equilibrium,  transformation of the  molecular motors and the geometrical confinement can give rise to very different performances according to the amplitude of the entropic barrier as well as on the phase shift between the two potentials. In the case of symmetric ratchet and channel,  we have show that $\Gamma$ has a non monotonous behavior for both processive and non processive motors, see Fig.~\ref{fig:simm-simm-no-force} leading to an optimal value of $\Delta S$ that minimizes the energy required to perform a single step. Such an optimal value of $\Delta S$ depends on the phase shift between the 
ratchet potential and the channel as well as on 
the processive nature of the motor. Surprisingly, the magnitude of the spatial confinement, quantified by  $\Delta S$ that minimizes $\Gamma$ is robust against variation in the ratchet potential depth $\Delta V$, see Fig.~\ref{fig:simm-simm-no-force-velVspot}. Finally, while for processive motors we have found an optimal value of $\Delta V$, for a given $\Delta S$, that minimizes $\Gamma$ (see Fig.~\ref{fig:simm-simm-no-force-velVspot}), for non processive motors $\Gamma$ decreases monotonously for increasing values of $\Delta V$. The inverse proportionality between the molecular motor average velocity $v$ and $\Gamma$, shown in Fig.~\ref{fig:simm-simm-no-force-vel}, underlines the mild dependence of the hopping dynamics on the confinement as opposed to the strong dependence experienced by $v$. 

For asymmetric  motors, which  rectify  even in the absence of  spatial confinement, we have found that  entropic barriers affect their overall performance.  The performance of both  processive and non processive motors  increases by properly tuning the amplitude and phase shift of the entropic barrier, as shown in Fig.~\ref{fig:asimm-simm-no-force}. Therefore, the geometrical constraints  reduce the molecular motor energy consumption to displace. Although in this case   $\Gamma$ always decreases when increasing  the binding motor potential, $\Delta V$, the dependence of $\Gamma$ on $\Delta V$, strongly depends on the processive nature of the motors. Specifically, we have seen that  for non processive motors the value of $\Delta S$ minimizing $\Gamma$ is quite insensitive to variations in $\Delta V$, while  the reverse holds  for processive motors. In particular 
we have found that while for smaller values of $\Delta V$ $\Gamma$ is minimum for $\Delta S=0$, when $\Delta V$ exceed a threshold the value of $\Delta S$ minimizing $\Gamma$ jumps to $\Delta S\neq 0$. The dependence of $v$ on $\Gamma$, as in the previous case, shows the mild dependence of he hopping dynamics on the confinement, see Fig.~\ref{fig:asimm-simm-no-force-vel}.

When motors are pulling against an applied force we can properly define an efficiency, $\eta$, as the ratio between the work performed over the energy consumed. 
We have seen that cooperative rectification allows symmetric molecular motors to  perform mechanical work against external forces. Moreover, we have found  that the efficiency increases with the external force at the expense of a reduction of the range of values of $\Delta S$ for which the motor can still step against the force. For larger forces motors are unable to step against and a confinement-dependent stall force can be defined. The presence of an external force dramatically changes the dependence of $\eta$  on $\Delta V$ for processive motors while non processive retain a behavior similar to that obtained for $f=0$, see Fig.~\ref{fig:simm-simm-no-force-velVspot}. In particular processive motors exhibit a $\Delta V$-dependent value of $\Delta S$ for which $\eta$ is 
maximized and the dependence of $\eta$ on $\Delta V$ identifies a $\Delta S$-dependent optimal value for which $\eta$ is maximum. On the contrary non processive motors show a $\Delta V$-independent optimal value of $\Delta S$ maximizing $\eta$ as we found in the case $f=0$. The dependence of $v$ on $\eta$, shown in Fig.~\ref{fig:simm-simm-force-vel} present the same behavior obtained in the previous cases, and the hopping dynamics is almost independent on the confinement.   

When the ratchet potential is asymmetric, i.e. motors rectify even for $\Delta S=0$, we found, as in the last case, that $\eta$ can be enhanced by properly tuning $\Delta S$ and $\Delta\phi$, as shown in Fig.~\ref{asimm-simm-force}. Moreover, both processive and non processive motors exhibit a maximum in $\eta$ for a $\Delta V$-independent value of $\Delta S$, see Fig.~\ref{fig:asimm-simm-force_velVSpot}. Interestingly, for processive motors such a behavior is different from both the case of asymmetric ratchet and $f=0$, see Fig.~\ref{fig:asimm-simm-no-force-velpot} and the case of symmetric ratchet and $f\neq 0$, see Fig.~\ref{fig:simm-simm-force_velVSpot} and reminds that of a symmetric ratchet potential in the absence of an external force, see Fig.~\ref{fig:simm-simm-no-force-velVspot}. 
Finally, the dependence of $v$ on $\eta$ is remarkably different from that obtained in all the previous cases. In fact, in the present case we found that $v$ and $\eta$ are not directly proportional, implying a more involved dependence of the hopping dynamics on the confinement. Such a different behavior underlines that the presence of an external force can strongly affect the inner dynamics of the motor therefore affecting the overall motor performance.  

J.M.R. and I.P. acknowledge  the Direcci\'on General de Investigaci\'on (Spain) and DURSI  for financial support
under projects  FIS\ 2011-22603 and 2014SGR922, respectively and financial support from {\sl Generalitat de Catalunya }under program {\sl Icrea Acad\`emia}.

\end{document}